\documentclass[final, 5p,times,twocolumn,authoryear]{elsarticle}

\usepackage{amsmath,amsfonts,amssymb}
\usepackage{algorithmic}
\usepackage{algorithm}
\usepackage{array}
\usepackage[caption=false,font=normalsize,labelfont=sf,textfont=sf]{subfig}
\usepackage{textcomp}
\usepackage{stfloats}
\usepackage[hidelinks]{hyperref}
\usepackage{verbatim}
\usepackage{graphicx}
\usepackage{cite}

\usepackage{tabularx}
\usepackage[most]{tcolorbox}
\usepackage{multicol}
\usepackage{mathpartir}
\usepackage{cleveref}
\usepackage{graphicx}  %
\usepackage{makecell}  %
\usepackage{placeins}
\usepackage{booktabs}
\usepackage{multirow}
\usepackage[hidelinks]{hyperref}
\usepackage{orcidlink}
\usepackage{amsthm}

\usepackage[dvipsnames]{xcolor}
\usepackage{adjustbox}
\usepackage{tikz}
\usetikzlibrary{patterns}
\usetikzlibrary{decorations.pathreplacing}
\usetikzlibrary{automata, arrows.meta, positioning}

\definecolor{BlueOutter}{RGB}{115, 141, 187}
\definecolor{BlueInner}{RGB}{221, 231, 250}
\definecolor{GreenOutter}{RGB}{140, 178, 111}
\definecolor{GreenInner}{RGB}{217, 231, 214}

\newcommand{\blackbubble}[1]{\tikz[baseline=(char.base)]{
            \node[draw=black, fill=black, thick, circle, inner sep=1pt] 
            (char) {\textcolor{white}{#1}}}}

\usepackage{xcolor}
\usepackage{pifont}

\usepackage{refcount}
\usepackage{listings}
\usepackage{turnstile}
\usepackage{caption}
\usepackage{fvextra}
\usepackage{newfloat}
\DeclareFloatingEnvironment[fileext=lol,placement={tbp},name=Listing]{listing}
\input{listings/pygments-xcode.tex}
\DefineVerbatimEnvironment{MintedVerbatim}{Verbatim}{}

\usepackage{xspace}

\usepackage{layout}

\newcommand{\tool}{\textsc{BASICS}\xspace}
\newcommand{\mss}{\texttt{MemStaCe}\xspace}

\newcommand{\LONG}[1]
{{}}

\newcommand{\DIFF}[1]{#1}

\newcommand\ib[1]{}
\newcommand\lf[1]{}

\newcommand{\mypara}[1]{\smallskip\noindent\textbf{{#1}.}}

\usepackage{ tikz, pgfplots, pgfplotstable, wrapfig, appendix }

\usetikzlibrary{ shapes.geometric, arrows, external, pgfplots.groupplots, matrix
  }

\pgfplotsset{compat=1.9}

\usepackage{mathtools,}

\DeclareMathAlphabet{\mathcal}{OMS}{cmsy}{m}{n}

\DeclareGraphicsExtensions{%
    .png,.PNG,%
    .pdf,.PDF,%
    .jpg,.mps,.jpeg,.jbig2,.jb2,.JPG,.JPEG,.JBIG2,.JB2}

\journal{Computers \& Security}

\makeatletter
\def\ps@pprintTitle{%
  \let\@oddhead\@empty
  \let\@evenhead\@empty
  \def\@oddfoot{\hbox to \textwidth{\footnotesize\itshape
    Accepted manuscript. Published version: \url{https://doi.org/10.1016/j.cose.2026.105165}\hfill}}%
  \let\@evenfoot\@oddfoot}
\makeatother

\begin{document}

\begin{frontmatter}

\title{\tool: Binary Analysis and Stack Integrity Checker System for Buffer Overflow Mitigation\tnoteref{am}}
\tnotetext[am]{This is the accepted manuscript of the article published in
\emph{Computers \& Security}, vol.~172, 105165 (2027),
\url{https://doi.org/10.1016/j.cose.2026.105165}.
\copyright~2026. This manuscript version is made available under the
CC~BY-NC-ND~4.0 license \url{https://creativecommons.org/licenses/by-nc-nd/4.0/}.}

\author[label1]{Luís Ferreirinha} \affiliation[label1]{organization={VUSec,
Vrije Universiteit Amsterdam}, addressline={De Boelelaan}, city={Amsterdam},
postcode={1081 HV}, state={Amsterdam}, country={The Netherlands}}

\author[label2]{Ibéria Medeiros\corref{cor1}} \affiliation[label2]{organization={LASIGE,
Faculty of Sciences of the University of Lisboa}, addressline={Campo Grande},
city={Lisboa}, postcode={1749}, state={Lisboa}, country={Portugal}}

\cortext[cor1]{Corresponding author: ivmedeiros@fc.ul.pt}

\begin{abstract}
  Linux-based systems rely on large ecosystems of packaged binaries, many of
  which are implemented in C/C++. While these languages provide low-level
  memory control and high performance, they are also highly susceptible to
  memory-safety vulnerabilities, particularly buffer overflows (BOs). Traditional vulnerability discovery techniques, such as static and dynamic analysis, often struggle with scalability and precision when applied directly to binary code, which can thereby keep programs vulnerable.
  This work introduces \tool, an approach and framework that combines model checking and bounded
  concolic execution to automatically verify security properties of stack memory
  in binary code, trampoline-based binary repair and crash-inducing inputs for bounded patch validation.
  \tool constructs a \textit{Memory State Space} -- \mss -- from the
  binary program's control-flow graph and bounded concolic execution of C
  function calls and loops. Security properties, defined in Linear Temporal
  Logic, model vulnerable stack behaviours and allow the approach to identify
  violations in \mss through counterexample traces. These vulnerabilities are
  then addressed using trampoline-based binary patching, and the resulting
  patches are validated using crash-inducing inputs extracted during concolic
  execution.
  We evaluated \tool on the Juliet C/C++
  and SARD benchmarks, as well as real open-source and Linux package binaries.
  \tool achieved 86.4\% precision and 87.4\% accuracy on SARD, and 86.1\%
  precision on Juliet, while also generating validated binary patches for
  supported vulnerable sinks. Compared with other evaluated
  tools, it provided the strongest overall balance between precision,
  analysis cost, and patching capability.
\end{abstract}

 \begin{keyword}
    Buffer Overflow Detection \sep Binary Code \sep Model Checking \sep Concolic
    Execution \sep Binary Patching \sep Software Security
 \end{keyword}

\end{frontmatter}

\section{Introduction}

Linux-based systems are deployed across a wide range of platforms, from
embedded devices and personal desktop computers to enterprise servers. The Linux kernel forms the core of these systems, while the surrounding user-space environment is composed of large ecosystems of packaged binaries, including
system utilities, long-running services, third-party applications, and embedded
user-space components~\citep{debian-policy-binary-packages,linux-kernel-userspace-api}. 
A significant fraction of this software is
implemented in C/C++, where low-level memory manipulation remains common.
Although this programming model gives developers precise control over data
layout, performance, and interaction with the operating system, it also exposes programs to memory-safety vulnerabilities~\citep{szekeres2013sok}. Stack-based buffer overflows (BOs) are a particularly important class of such vulnerabilities: an out-of-bounds write to a
stack object can corrupt adjacent local variables, saved frame pointers, return
addresses, or other control-relevant data, potentially enabling crashes,
control-flow hijacking, or arbitrary code execution~\citep{cwe121}.

Despite widespread compiler and operating-system mitigations, including
stack canaries, address-space layout randomisation, non-executable stacks, and control flow hardening, BOs have not disappeared from native software deployed~\citep{on_the_effectiveness_address_randomization, cfi, Stackguard,cyberdive:25} and remain ranked by CWE as one of the most dangerous vulnerabilities~\citep{CWETop25}.
In practice, hardening is not uniform across all binaries, and some protections depend on compiler options, linker options, source-level patterns, or deployment-specific build configurations~\citep{openssf-compiler-hardening}. Consequently, security analysts often need to reason directly about binaries: they must detect whether a compiled program contains a vulnerable stack write, identify where the vulnerability occurs, and, when source-level repair is unavailable or impractical, apply a mitigation directly to the binary.

Detecting vulnerabilities such as BOs has been a long-standing problem, with a
plethora of different tools and methods developed~\citep{delta, CorCA,
BOFSanitizer, russell2018automated, liu_csod_2019, BofAEG}. These tools employ various analysis methods, but often follow one of the following approaches: Static Analysis, Dynamic Analysis, or a hybrid of both. Static analysis analyses a program's code without executing it, achieving high code coverage at the cost
of increased false positives~\citep{binary_code_is_not_easy}. In contrast,
dynamic analysis techniques execute the code of a
program~\citep{the_concept_of_dynamic_analysis}, allowing for more accurate
vulnerability detection, but at the cost of code coverage. Hybrid approaches
have been proposed to mitigate these limitations~\citep{CorCA,
static_dynamic_detecting_vulnerabilities,
vulnerability_detection_static_dynamic_analysis, vadayath:22}.
Although improvements were achieved, these techniques often struggle with
scalability and precision when applied to binary code, still leaving programs
vulnerable.  

In addition to this detection deficit, these vulnerabilities, when discovered,
are generally reported to developers, whose job is to remove them. However, developers cannot always provide an immediate patch, which often results in a
prolonged vulnerability life of up to 12 months in some
cases~\citep{patching_vulnerabilities}, thus keeping an open window for
attackers to continue exploiting the systems that contain them.
Research on this issue has led to the development of methods that automatically
patch vulnerabilities. Although most approaches focus on source
code~\citep{CorCA, BOFSanitizer, autopag_automated_software_patch_generation},
recent work has expanded to binary-level patching. Tools such as
E9Patch~\citep{e9patch} allow users to replace sections of a binary with custom
patches, and other works aim to completely automate this process
\citep{duan2019automating, patching_vulnerabilities}.
However, to effectively remove these vulnerabilities, one must first identify
their location within the binary, a challenging task with disassembled programs that offer little insight into the higher-level logic present in the source
code. Then, as a second step, the validation of such patches involves checking
that they effectively remove the target vulnerabilities without introducing new
ones and compromising the program's correct functioning. 
Therefore, a significant gap can then be identified in the current landscape of
vulnerability mitigation tools: \textit{the absence of a scalable, accurate, and
automated solution to identify and patch vulnerabilities in binary programs}.

This paper presents a novel approach to automatically detect and mitigate
BOs in compiled C binaries. 
We call this approach \textit{Binary Analysis and Stack Integrity Checker System} (\tool).
\tool combines static binary analysis,
bounded symbolic execution, and model checking. Static analysis recovers the
binary-level control-flow structure, bounded symbolic execution refines the
stack effects of calls and bounded loops, and model checking verifies Linear Temporal Logic (LTL)-specified stack-security properties over the resulting stack memory model.
When a violation is found, a counterexample is produced and used to localise the vulnerable sink and guide trampoline-based binary patching, followed by bounded validation of the patched binary.

The most relevant works on software model checking in the literature tend to
focus on verifying models of distributed systems~\citep{spin_paper}, or on
bounded verification of source-level C programs~\citep{cbmc}. Although these do
not directly detect vulnerabilities, they can be used to discover design flaws
in programs that might lead to a vulnerability. Moreover, the use of symbolic
and concolic methods have been widely used to improve vulnerability discovery, often in
combination with fuzzing
execution~\citep{qsym_concolic_execution_hybrid_fuzzing,
symsan_efficient_concolic_execution_data_flow_analysis,
injection_vulns_concolic}. Such approaches typically search for crashing inputs
or path witnesses rather than constructing a temporal memory model for checking
explicit stack-integrity properties. Finally, automatic program repair has
mainly focused on source-level patches~\citep{monperrus_repair}, whereas
binary-level repair remains more challenging because source-level structure and
type information are unavailable. The proposed approach addresses this gap by constructing a binary-level memory-state model, checking LTL stack-security properties over that model, and applying validated trampoline-based patches directly to the binary.

The approach models the stack memory of the binary program as a \textit{Memory State Space}~(\mss). It constructs this state space from the binary control-flow graph and refines it with bounded symbolic execution, which is used to approximate the stack effects of selected C library calls, indirect data-dependent writes, and bounded loop executions. In \mss, each transition between memory states corresponds to an operation in the stack memory (e.g., pushing a register to the stack). Afterwards, \mss is used to check whether security properties are violated, thus flagging the potential existence of BOs and generating a counterexample trace with the instructions involved in each violation. Security properties model the correct usage of the stack memory space of functions related to BO vulnerabilities (e.g., \texttt{strcpy}). These are
defined in Linear Temporal Logic (LTL)~\citep{handbook_model_checking} and
translated into Büchi-Automaton before being used by the model checker in the BO
discovery task. Traces are submitted to a reverse-flow analysis to pinpoint the
location of the vulnerability sink (a function like \texttt{strcpy} that is
sensitive to malicious inputs), and then patched with binary fixes (small binary
programs containing the correct code that fix the vulnerability) to remove the
found BOs by using a trampoline-based binary patching method. Lastly, the
approach checks the effectiveness of the patches using vulnerable concolic
inputs derived from the symbolic execution process, alongside benign and
boundary inputs.

The paper also presents \tool's implementation for detecting and removing BOs, with its effectiveness validation. We evaluated \tool based on three
different criteria. First, we assessed its ability to detect BO vulnerabilities
in binaries using the Juliet C/C++ test suite and a subset of the NIST
SARD~\citep{nist-sard} dataset and compared its effectiveness against other tools, achieving the strongest overall balance between precision, analysis cost, and patching capabilities.
Second, we evaluated how \tool is effective in mitigating BOs with the
SARD subset. Finally, we evaluated the scalability of \tool, by performing a
sweep of 292 binaries from Linux repositories and 3 large open-source projects.

The main contributions of the paper are: 
(1)  a novel binary-level approach that combines static analysis, bounded symbolic execution, model checking, and trampoline-based patching to detect, localise, and mitigate stack-based BO vulnerabilities; 
(2) a Memory State Space (\mss) abstraction for representing
binary-level stack behaviour, constructed from control-flow recovery,
stack-memory modelling, and bounded symbolic exploration of calls and loops;
(3) a framework to define security properties to model the correct usage of \mss, resorting to Linear Temporal Logic (LTL); 
(4) a trampoline-based binary patching method, including patch templates, to remove BOs and validate the correction;
(5) the \tool tool (available at~\citep{basics_github_a}) that implements the
main approach, including the aforementioned methods; 
(6) an evaluation that demonstrates the ability of \tool in the detection and removal of vulnerabilities. 

The remaining of the paper is organised as follows. Section~\ref{seq:background}
details the background that supports our approach, Sections~\ref{sec:overview}
to \ref{sec:implementation} present in detail the approach itself, how it works,
and its implementation, Section~\ref{sec:evaluation} evaluates the approach, Section~\ref{sec:lim} its limitations, and
Sections~\ref{sec:extension}, \ref{sec:rw} and \ref{seq:conclusions} present,
respectively, how \tool can be extended to other vulnerabilities, related work,
and conclusions.

\section{Concepts}
\label{seq:background}

This section presents the key concepts necessary for a deeper understanding of
our approach. Firstly, it provides an overview of a BO vulnerability, the type
of vulnerability we focus on throughout the paper's explanations, and then the
techniques behind our approach -- model checking, linear temporal logic (LTL), and concolic execution. 

\subsection{Stack Buffer Overflows}

Languages such as C/C++ do not have built-in safeguards to prevent the
programmer from accessing memory outside the bounds of buffers (a continuous,
fixed-sized region of the stack memory used to store local variables, function
parameters, and function return addresses \citep{szekeres2013sok}). As a result,
out-of-bounds memory operations can occur, such as writing beyond the upper
boundary of a buffer and overwriting function return addresses, which are
essential to maintain the intended control flow of the program. These lead to a
weakness known as a buffer overflow (BO) vulnerability.
Depending on the nature of the overflow, they might be benign or cause a program
to crash outright, but in some cases an attacker can influence the buffer's contents and exploit it, redirecting execution to malicious code.

\begin{listing}[htb]
   \begingroup\fvset{highlightcolor=gray!20,frame=lines, numbers=left, numbersep=-5pt, xleftmargin=0pt, fontsize=\footnotesize, framesep=2mm}\begin{MintedVerbatim}[commandchars=\\\{\}]
\PYG{+w}{  }\PYG{k+kt}{void}\PYG{+w}{ }\PYG{n+nf}{copy}\PYG{p}{(}\PYG{k+kt}{char}\PYG{+w}{ }\PYG{o}{*}\PYG{n}{str}\PYG{p}{)}\PYG{+w}{ }\PYG{p}{\PYGZob{}}
\PYG{+w}{      }\PYG{k+kt}{char}\PYG{+w}{ }\PYG{n}{buffer\PYGZus{}2}\PYG{p}{[}\PYG{l+m+mi}{16}\PYG{p}{];}
\PYG{+w}{      }\PYG{n}{strcpy}\PYG{p}{(}\PYG{n}{buffer\PYGZus{}2}\PYG{p}{,}\PYG{+w}{ }\PYG{n}{str}\PYG{p}{);}
\PYG{+w}{  }\PYG{p}{\PYGZcb{}}\PYG{+w}{ }
\PYG{+w}{  }
\PYG{+w}{  }\PYG{k+kt}{void}\PYG{+w}{ }\PYG{n+nf}{main}\PYG{p}{()}\PYG{+w}{ }\PYG{p}{\PYGZob{}}
\PYG{+w}{      }\PYG{k+kt}{char}\PYG{+w}{ }\PYG{n}{buffer\PYGZus{}1}\PYG{p}{[}\PYG{l+m+mi}{256}\PYG{p}{];}
\PYG{+w}{      }
\PYG{+w}{      }\PYG{k}{for}\PYG{+w}{ }\PYG{p}{(}\PYG{k+kt}{int}\PYG{+w}{ }\PYG{n}{i}\PYG{+w}{ }\PYG{o}{=}\PYG{+w}{ }\PYG{l+m+mi}{0}\PYG{p}{;}\PYG{+w}{ }\PYG{n}{i}\PYG{+w}{ }\PYG{o}{\PYGZlt{}}\PYG{+w}{ }\PYG{l+m+mi}{255}\PYG{p}{;}\PYG{+w}{ }\PYG{n}{i}\PYG{o}{++}\PYG{p}{)}\PYG{+w}{ }
\PYG{+w}{          }\PYG{n}{buffer\PYGZus{}1}\PYG{p}{[}\PYG{n}{i}\PYG{p}{]}\PYG{+w}{ }\PYG{o}{=}\PYG{+w}{ }\PYG{l+s+sc}{\PYGZsq{}x\PYGZsq{}}\PYG{p}{;}
\PYG{+w}{      }\PYG{n}{copy}\PYG{p}{(}\PYG{n}{buffer\PYGZus{}1}\PYG{p}{);}
\PYG{+w}{  }\PYG{p}{\PYGZcb{}}
\end{MintedVerbatim}
\endgroup
      \vspace{-3mm}    
   \caption{Stack buffer overflow vulnerability example in C.}
   \label{stack_overflow_example}
\end{listing}

The code in Listing \ref{stack_overflow_example} demonstrates a standard example
of a BO vulnerability. In this code snippet, a 256-byte buffer
(\texttt{buffer\_1} on line 7) is allocated and filled with the character
\texttt{x} (lines 9--10). Subsequently, the \texttt{copy} function (line 11) is
invoked with this buffer as a parameter to copy its content to a 16-byte buffer
(\texttt{buffer\_2} on line 2), calling the \texttt{strcpy} function from the C
standard library. The \texttt{strcpy} function is considered dangerous as it
does not take into account the size of the destination buffer when copying
contents between buffers. Because \texttt{buffer\_2} is not large enough to
accommodate the data from \texttt{buffer\_1}, this operation results in a BO,
where excess data spills into the adjacent memory space.

By compiling the previous code to Assembly x86-64, we gain the ability to
analyse the inner workings of the \texttt{copy} function and its interactions
with memory.
The code in Listing~\ref{copy_function_asm} shows this assembly code of the
stack frame for the \texttt{copy} function call. 
Initially, the register base pointer (\texttt{RBP}) of the previous function
(\texttt{main}) is preserved on the stack with the instruction \texttt{push
rbp}, leaving the register \texttt{RBP} free to receive the current value of the
register stack pointer (\texttt{RSP}), denoted by the instruction \texttt{mov
rbp, rsp}, and thus stating the stack frame base for the \texttt{copy} function. Afterwards, \texttt{RSP} decreases by \texttt{32} bytes with
\texttt{sub rsp, 32}, allocating space for the local variables of the
\texttt{copy} function in its stack frame. Within this space, an 8-byte pointer
to \texttt{buffer\_1} is stored at the address \texttt{RBP-24} (line 4),
provided by the \texttt{RDI} register when the function \texttt{copy} is called.
Furthermore, the address of the \texttt{buffer\_2} located at the
\texttt{RBP-16} (line 6) is assigned to the \texttt{RAX} register. Then, lines 7
and 8 prepare the arguments for invoking the \texttt{strcpy} function. When
\texttt{strcpy} is invoked, it is instructed to copy data from the location
pointed to by \texttt{RDI} (line 8, which currently is \texttt{buffer\_1}) to
the space starting at the address \texttt{RBP-16} (represented by \texttt{RSI} and
indicating where \texttt{buffer\_2} starts). Since \texttt{buffer\_1} contains
256 bytes of data, it far exceeds the 16-byte capacity of \texttt{buffer\_2}.
Consequently, excess data from \texttt{buffer\_1} overflows and corrupts the
adjacent memory space beyond \texttt{buffer\_2}, thus overflowing the control
data stored in the stack, namely \texttt{RBP} and \texttt{RIP} (register
instruction pointer), the last being the pointer of the caller function (line
11 of Listing \ref{stack_overflow_example}). 

\begin{listing}[t]
   \begingroup\fvset{highlightcolor=gray!20,frame=lines, numbers=left, numbersep=-5pt, xleftmargin=0pt, fontsize=\footnotesize, framesep=2mm}\begin{MintedVerbatim}[commandchars=\\\{\}]
\PYG{+w}{   }\PYG{n+nf}{push}\PYG{+w}{    }\PYG{n+nb}{rbp}
\PYG{+w}{   }\PYG{n+nf}{mov}\PYG{+w}{     }\PYG{n+nb}{rbp}\PYG{p}{,}\PYG{+w}{ }\PYG{n+nb}{rsp}
\PYG{+w}{   }\PYG{n+nf}{sub}\PYG{+w}{     }\PYG{n+nb}{rsp}\PYG{p}{,}\PYG{+w}{ }\PYG{l+m+mi}{32}
\PYG{+w}{   }\PYG{n+nf}{mov}\PYG{+w}{     }\PYG{k+kt}{QWORD}\PYG{+w}{ }\PYG{n+nv}{PTR}\PYG{+w}{ }\PYG{p}{[}\PYG{n+nb}{rbp}\PYG{o}{\PYGZhy{}}\PYG{l+m+mi}{24}\PYG{p}{],}\PYG{+w}{ }\PYG{n+nb}{rdi}
\PYG{+w}{   }\PYG{n+nf}{mov}\PYG{+w}{     }\PYG{n+nb}{rdx}\PYG{p}{,}\PYG{+w}{ }\PYG{k+kt}{QWORD}\PYG{+w}{ }\PYG{n+nv}{PTR}\PYG{+w}{ }\PYG{p}{[}\PYG{n+nb}{rbp}\PYG{o}{\PYGZhy{}}\PYG{l+m+mi}{24}\PYG{p}{]}
\PYG{+w}{   }\PYG{n+nf}{lea}\PYG{+w}{     }\PYG{n+nb}{rax}\PYG{p}{,}\PYG{+w}{ }\PYG{p}{[}\PYG{n+nb}{rbp}\PYG{o}{\PYGZhy{}}\PYG{l+m+mi}{16}\PYG{p}{]}
\PYG{+w}{   }\PYG{n+nf}{mov}\PYG{+w}{     }\PYG{n+nb}{rsi}\PYG{p}{,}\PYG{+w}{ }\PYG{n+nb}{rdx}
\PYG{+w}{   }\PYG{n+nf}{mov}\PYG{+w}{     }\PYG{n+nb}{rdi}\PYG{p}{,}\PYG{+w}{ }\PYG{n+nb}{rax}
\PYG{+w}{   }\PYG{n+nf}{call}\PYG{+w}{    }\PYG{n+nv}{strcpy}
\PYG{+w}{   }\PYG{n+nf}{nop}
\PYG{+w}{   }\PYG{n+nf}{leave}
\PYG{+w}{   }\PYG{n+nf}{ret}
\end{MintedVerbatim}
\endgroup
    \vspace{-3mm}
    \caption{\texttt{copy} function's x86-64 assembly code}
   \label{copy_function_asm}
\end{listing}

\subsection{Model Checking}

Model Checking is a computational technique used to analyse the behaviours of
dynamic systems, which are represented as state transition
systems~\citep{handbook_model_checking}. This model retains the system's
essential properties and allows for the verification of a system's design when
its complete implementation is too complex to verify directly.

A model checker can be described as being composed of three main components
\citep{handbook_model_checking}: \textbf{Model}: A finite state-transition graph
that provides adequate formalism for the description of the system, generally
designated as a Kripke Structure, denoted as $K$; \textbf{Specification}: The
system's desired properties are expressed as temporal logic formulas $\varphi$,
which provide a framework for specifying the correctness criteria of state
transitions, i.e., the system's behaviour; \textbf{Algorithm}: A computational
method used to determine whether the state transition model follows the
specifications outlined in the temporal logic formulae.

Together, these components facilitate the model checking process. The model
checker employs a decision procedure to determine whether $K \models \varphi$
holds, i.e., if the Kripke structure $K$ satisfies the property $\varphi$. Should
$K$ not satisfy $\varphi$ (expressed as $K \not\models \varphi$), the model
checker provides a counterexample (i.e., a correctness proof), demonstrating how
the security property $\varphi$ is violated within the structure $K$.

\vspace{-1mm}
\subsection{Linear Temporal Logic}

Temporal logic is used to reason about the way the world changes over time. In
the context of software, it is used in the specification and description of
systems by describing the evolution of states of a program, which gives rise to
descriptions of executions~\citep{handbook_model_checking}. 

Linear Temporal Logic (LTL), as the name implies, follows the linear-time
progression view. In addition to the operators present in propositional logic,
this logic provides temporal operators that connect different stages of the
computations and represent dependencies and relations between them. LTL formulas
are constructed using normal Boolean operators ($\lnot,\,\lor,\,\land$) and the
temporal operators \textit{next}, \textit{previous}, \textit{until} and
\textit{since}.
These operators can then be used to define temporal abbreviations, with the
most commonly used operators in LTL formulae being: $\Diamond \varphi$ (eventually),
$\square \psi$ (always), $\varphi \mathcal{W} \psi$ (Weak-Until), $\varphi
\mathcal{R} \psi$ (Release)~\citep{handbook_model_checking}.

To facilitate the model checking process, an LTL formula can be converted to an
$\omega$-automaton, thus enabling the formalisation of the model
checking problem as a search for accepted runs on the synchronous product of the State Space and a $\omega$-automaton~\citep{simple_verification_ltl}.

\subsection{Concolic Execution}

Concolic Execution is a method that combines symbolic and concrete execution,
meaning that symbolic and concrete values are used for inputs, and the program
is executed both symbolically and concretely. The concrete execution part
constitutes the normal execution of the program, while the symbolic execution
collects symbolic constraints over the symbolic input values at each branch
point encountered along the concrete execution path \citep{concolic_testing}.
The process starts with the execution of the program on a set of initial inputs.
As the program runs, it collects constraints on the inputs from conditional
statements encountered along the execution path. These constraints are then used
to generate a symbolic representation of the program execution, capturing the
relationships between inputs and the program's behaviour. To solve these
constraints and determine whether a path is executable, Satisfiability Modulo
Theories (SMT) Solvers~\citep{de2008z3} are used.

\begin{figure*}[!ht]
    \makebox[\textwidth][c]{\includegraphics[width=\textwidth]{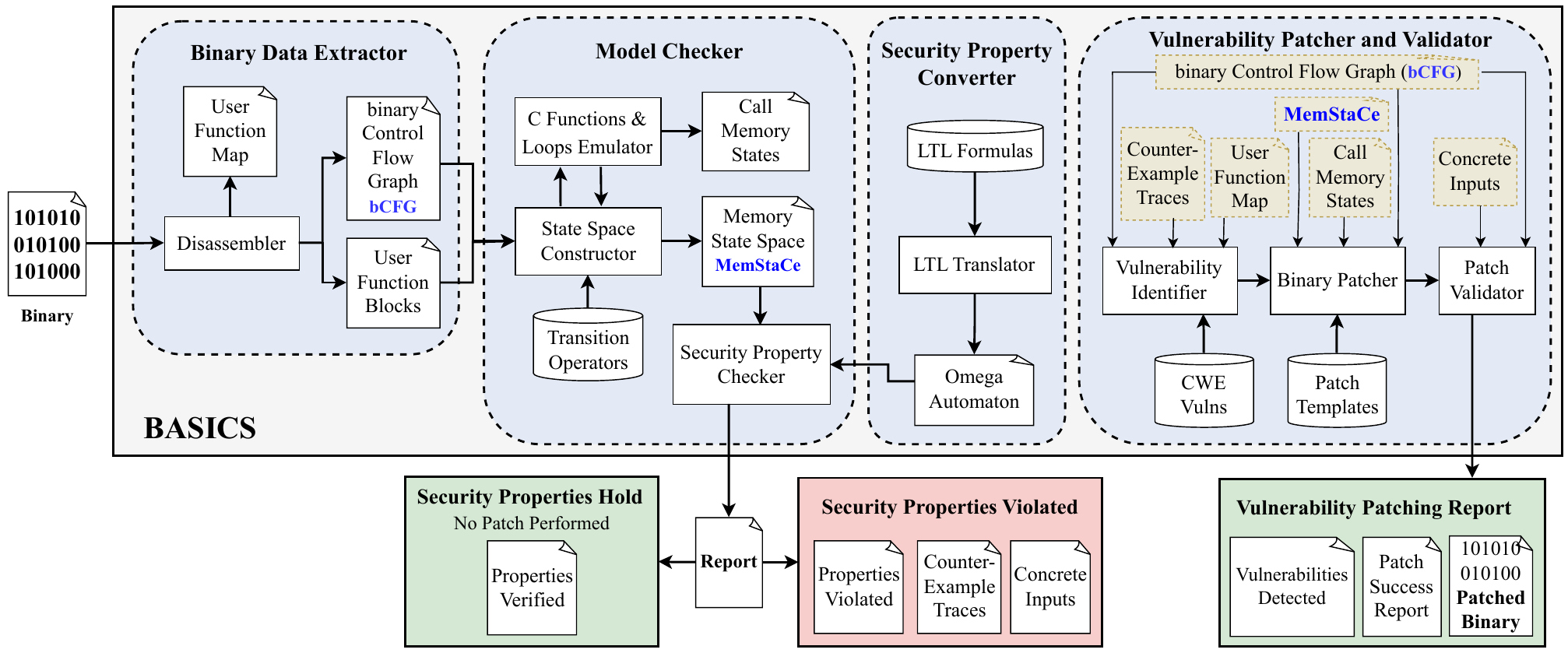}}
    \vspace{-5mm}
    \caption{Overview of the proposed \tool's approach.}
    \label{fig:arch}
    \vspace{-3mm}
\end{figure*}

\section{\tool Approach Overview}
\label{sec:overview}

\tool's approach analyses binary programs to detect stack-buffer-overflow vulnerabilities
and repair them, including correctness validation.
The approach first constructs a
\emph{Memory State Space} (\mss), an abstract transition system that records
program locations together with stack configurations. The \mss captures direct
stack effects of assembly instructions and incorporates summary transitions for
external C library calls and loops, whose effects are recovered through static
argument recovery and bounded concolic execution.

Security properties are expressed in Linear Temporal Logic (LTL) over
observations of \mss states, such as stack-frame metadata, byte states, recovered
buffer boundaries, and the previous transition. These properties are translated to automata and checked over the \mss. When a property is violated, \tool emits a counterexample trace that identifies the sequence of binary instructions and abstract memory actions that lead to the unsafe stack state. This trace is used to classify the finding and, when the violation is linked to a supported call sink, to conduct binary patch generation.

For patchable findings, \tool selects a patch template for the vulnerable sink,
instantiates it using recovered call-state information such as argument
registers, destination-buffer bounds, stack-canary metadata, and continuation
addresses, and installs the payload into the binary using E9Patch. The patched
binary is then validated under bounded executions using malicious, benign, and
boundary inputs. This validation provides evidence that the observed vulnerable
behaviour is remediated and that tested benign and boundary behaviours are
preserved.

As shown in \Cref{fig:arch}, the approach consists of four modules:
\emph{Binary Data Extractor}, \emph{Model Checker}, \emph{Security Property
Converter}, and \emph{Vulnerability Patcher and Validator}. We provide a quick
overview of these next.

\mypara{Binary Data Extractor}
This module is responsible for disassembling a binary program, extracting its binary Control Flow Graph (bCFG), the User Functions Definition it contains, including function names and starting addresses, and the User Function Address Map that will serve as a reference for later vulnerability detection and patching. Similarly to a CFG extracted from the source code, the bCFG contains the basic blocks representing assembly instructions and the branches derived from
control-transfer instructions (e.g., \texttt{jump}). We derive user-function
information by iterating over the basic blocks of bCFG and identifying the entry address of each function, i.e., the start address of its defining block.

\mypara{Model Checker}
This module constructs the \mss by traversing the recovered bCFG and translating
stack-relevant instructions into abstract memory actions. It models direct stack
updates, such as writes, pushes, pops, frame extensions, frame activation, buffer
metadata, and canary metadata. It also incorporates summary transitions for
external calls and loops. These summaries are computed using recovered call
arguments and bounded symbolic execution, and may produce concrete inputs for
later validation. The model checker then evaluates the automata generated from
the LTL properties over the \mss and emits one of two reports: 
\emph{Security Properties Violated}, containing the list of the
violated properties, the corresponding counterexample traces, and the concreted
inputs; or \emph{Security Properties Hold}, containing the properties satisfied
by the explored \mss.

\mypara{Security Property Converter}
This module translates LTL security properties into automata used during model
checking. The properties specify unsafe stack behaviours, such as modification of
control-sensitive bytes, use of unsafe library calls, and writes that cross
recovered stack-buffer boundaries. The resulting automata are checked against
the property-specific Kripke structure derived from the \mss.

\mypara{Vulnerability Patcher and Validator}
This module consumes violated properties and their counterexample traces. It
first identifies whether the finding is associated with a supported patchable
sink, currently external-library call sites such as \texttt{strcpy},
\texttt{gets}, \texttt{sprintf}, \texttt{memcpy}, and related functions. If the
finding is patchable, the module selects and instantiates a patch template,
installs the generated payload into the binary using E9Patch, and records patch
installation evidence. Finally, it validates the patched binary under bounded
malicious, benign, and boundary executions, reporting remediation,
preservation, failure, or inconclusive outcomes.

The following sections describe each stage in detail. \Cref{sec:model_checker}
defines the construction of \mss, \Cref{sec:secProper} explains how LTL
properties are checked and how violations become vulnerability reports, and
\Cref{sec:patcher} presents patch generation and bounded validation. The
implementation of these modules in \tool is described in
\Cref{sec:implementation}.

\section{Constructing the Memory State Space}
\label{sec:model_checker}

This section presents our proposed memory state space for binaries (\mss), starting with the formal definition of its components, and culminating in its definition, followed by the construction of external calls and loops.

\subsection{Formal Definition of MemStaCe}

The model checker constructs the \textit{Memory State Space} (\mss) of a binary
program and explores it to detect stack-memory states that may violate security
properties related to buffer overflows. Formally, \mss is constructed in several
stages. First, \tool recovers a binary control-flow graph (bCFG) that captures the
possible execution paths of the program. Second, it defines an abstract
representation of stack memory consisting of stack frames, buffer metadata, and
security-relevant byte states. Third, assembly instructions are mapped to
abstract memory actions that model stack updates independently of the underlying
instruction set. Fourth, operational semantics are defined for these actions,
describing how they transform abstract stack configurations. Finally, the
resulting stack transitions are combined with control-flow transitions to
produce \mss, a transition system whose states represent both
the current program location and the corresponding abstract stack state.

\mypara{Binary Control Flow Graph}
For a binary program $P$, let
\[
    bCFG_P = (V,E,\rho_0,\mathsf{inst})
\]
be the binary control-flow graph recovered by \tool. The set $V$ contains
program locations, $E \subseteq V \times V$ is the control-flow edge relation,
and $\rho_0 \in V$ is the entry location. If $(\rho,\rho') \in E$, execution may
flow from $\rho$ to $\rho'$. The map
\[
    \mathsf{inst} : V \to \mathtt{Instr}
\]
assigns an instruction to each program location.
When the recovered bCFG contains basic-block nodes, we assume the graph is
normalized so that each node contains at most one memory-relevant instruction.
Equivalently, a basic block with multiple memory-relevant instructions is split
into a sequence of program locations.

\mypara{Stack Model}
Let \(\mathsf{Stack}\) denote the set of all abstract stack configurations. A
stack configuration \(K \in \mathsf{Stack}\) is an ordered sequence of stack
frames:
\[
    K = F_0 \cdot F_1 \cdots F_n,
    \qquad
    \operatorname{top}(K) = F_n 
\]
A single execution path has one current stack configuration at each program
location. \mss may contain many such configurations because different
control-flow paths can produce different stacks. The order is significant: $F_n$
is the currently active frame, while $F_0,\ldots,F_{n-1}$ are caller frames. 

Before we define a frame, we introduce buffer descriptors and byte-states.
For a given frame with $p$ recorded stack buffers, the buffer metadata is a set
of buffer descriptors $\sigma$
\[
  \Sigma : \{\sigma_0, \dots, \sigma_{p-1}\}
\]
A buffer descriptor is a pair $\sigma = (o,n)$,
where $o$ is the starting offset (i.e., $\operatorname{start}(\sigma)=o$) and
$n$ is the size of the buffer; so $\operatorname{end}(\sigma)=o+n-1$. 

Additionally, for a frame with $m$ modelled bytes, the byte-state array is a
finite map
\[
    B : \{0,\ldots,m-1\} \to \mathit{BState}, 
\]
with possible byte states being
\[
bs \in \mathit{BState}
  ::= \mathrm{Free}
   \mid \mathrm{Occupied}
   \mid \mathrm{Critical}
   \mid \mathrm{Modified}
\]
The state $\mathrm{Free}$ denotes reserved but uninitialized stack space;
$\mathrm{Occupied}$ denotes ordinary initialized stack data; $\mathrm{Critical}$
denotes control or security-sensitive data, such as saved return addresses or
frame pointers; and $\mathrm{Modified}$ denotes a byte that has been overwritten
after initialization.

If a stack canary is detected for a frame, its presence is recorded by the
Boolean metadata field \(\gamma\):
\[
    \gamma \in \mathbb{B} ::= \mathsf{true} \mid \mathsf{false}.
\]
Finally, we define a frame as a tuple that records a function label, a
byte-state array, buffer metadata, and canary metadata:
\[
    F = (\ell,B,\Sigma,\gamma).
\]

\mypara{Abstract Memory Actions}
\mss does not give semantics directly to assembly instructions. Instead, each
instruction is first translated into an abstract memory action defined in
\Cref{fig:abstract-memory-actions}.

\begin{figure}[b]
\centering
\small
\[
\begin{array}{rcl}
a \in \mathit{Act}
& ::= &
\mathrm{ActivateFrame(\ell)}
\mid \mathrm{Write}(o,w,bt)
\mid \mathrm{Extend}(n)
\\[0.35em]
&&
\mid \mathrm{Push}(n,bt)
\mid \mathrm{Pop}(n)
\mid \mathrm{Buf}(\sigma)
\mid \mathrm{ExtCall}(c,\beta)
\\[0.35em]
&&
\mid \mathrm{Loop}(L,\beta)
\mid \mathrm{Canary}
\mid \epsilon .
\end{array}
\]
\vspace{-3mm}
\hrule
\vspace{2mm}
\caption{Abstract memory actions.}
\label{fig:abstract-memory-actions}
\end{figure}

\begin{figure}[htb]
\centering
\small
\[
\begin{gathered}
\mathit{bt} \in \mathit{BTrans}
  ::= \mathrm{nRWrite} \mid \mathrm{RWrite}
\\[0.8em]
\textit{Byte transition.}
\qquad
\boxed{\delta_{bt} : \mathit{BState} \to \mathit{BState}}
\\[0.8em]
\begin{aligned}
\delta_{\mathrm{nRWrite}}(q)
&\triangleq
\begin{cases}
\mathrm{Occupied}
  & q = \mathrm{Free},\\
\mathrm{Modified}
  & q \in \{\mathrm{Occupied},\mathrm{Critical},\mathrm{Modified}\},
\end{cases}
\\[1.2em]
\delta_{\mathrm{RWrite}}(q)
&\triangleq
\begin{cases}
\mathrm{Critical}
  & q = \mathrm{Free},\\
\mathrm{Modified}
  & q \in \{\mathrm{Occupied},\mathrm{Critical},\mathrm{Modified}\}.
\end{cases}
\end{aligned}
\\[1em]
\textit{Byte-array update.}
\qquad
\boxed{B[o,w \leftarrow bt]}
\\[0.8em]
B[o,w \leftarrow bt] \triangleq B'
\quad\text{where}\quad
\operatorname{dom}(B')=\operatorname{dom}(B)
\\[0.4em]
\text{and}\qquad
\forall i \in \operatorname{dom}(B).\;
B'[i] =
\begin{cases}
\delta_{bt}(B[i]) & o \le i < o+w,\\
B[i] & \text{otherwise.}
\end{cases}
\end{gathered}
\]
\hrule
\vspace{2mm}
\caption{Byte transitions and byte-array update semantics.}
\label{fig:byte-write-semantics}
\end{figure}
The action \(\mathrm{ActivateFrame}(\ell)\) makes the frame associated with
function label \(\ell\) the active frame. If no such frame exists in the current
stack configuration, it creates a fresh frame for \(\ell\). Otherwise, it moves
the existing frame for \(\ell\) to the top of the stack. The action
$\mathrm{Write}(o,w,bt)$ updates $w$ bytes starting at offset $o$, according to
the specified \emph{byte-transition}, $bt$. These are defined
in~\Cref{fig:byte-write-semantics}, and will be discussed later alongside the
semantics of the actions. The actions $\mathrm{Extend}(n)$,
$\mathrm{Push}(n,bt)$, and $\mathrm{Pop}(n)$, respectively, allocate, append,
and remove bytes from the current frame. The action $\mathrm{Buf}(\sigma)$
records buffer metadata. Calls and loops are represented by
$\mathrm{ExtCall}(c,\beta)$ and $\mathrm{Loop}(L,\beta)$, where \(c\) identifies
the program location \(\rho\) at which the external function call occurs, \(L\)
identifies the program location \(\rho\) of the loop header or entry point, and
\(\beta\) represents the bounds used during bounded symbolic execution. 
The action $\mathrm{Canary}$ denotes the presence of a canary in the stack frame, and the action \(\epsilon\) symbolises an instruction with no relevant stack-memory effect.

We write
\[
    \alpha : V \to \mathit{Act}
\]
for the labeling function that maps each program location to its corresponding
abstract memory action:
\[
    \alpha(\rho) \triangleq \mathrm{MemAct}(\mathsf{inst}(\rho)),
\]
where $\mathrm{MemAct} : \mathtt{Instr} \to \mathit{Act}$ abstracts concrete
instructions into \mss actions.

\mypara{Semantics}
The semantics of memory actions are defined in four layers. First, byte
transitions describe how individual byte states evolve when a write occurs.
Second, stack operations define changes to the active stack-frame context.
Third, frame operations lift byte transitions to individual stack frames.
Finally, action semantics lifts stack and frame operations to complete stack
configurations and defines how \mss transitions follow the bCFG.

\Cref{fig:byte-write-semantics} defines the byte-level semantics. The transition
type \(bt\) distinguishes ordinary non-risky stack writes, represented by
\(\mathrm{nRWrite}\), from writes that initialise control-sensitive stack data,
represented by \(\mathrm{RWrite}\). A non-risky write changes a
\(\mathrm{Free}\) byte into \(\mathrm{Occupied}\), while a risky write changes a
\(\mathrm{Free}\) byte into \(\mathrm{Critical}\). In both cases, writing to a
byte that was already \(\mathrm{Occupied}\), \(\mathrm{Critical}\), or
\(\mathrm{Modified}\) changes it to \(\mathrm{Modified}\). Thus, the
\(\mathrm{Modified}\) state records that a previously initialised byte has been written again.

The byte-array update \(B[o,w\leftarrow bt]\) applies the corresponding byte
transition to all bytes in the interval \([o,o+w)\), while leaving all other
bytes unchanged. The domain of the byte array is preserved by the update; the
validity of the accessed interval is enforced by the frame-level write rule.

\Cref{fig:frame-semantics} defines stack-level and frame-level operations. The
stack-level operation \(\operatorname{activate}(K,\ell)\) changes which frame is active in the stack configuration \(K\). If \(K\) does not contain a frame
labelled \(\ell\), the rule \textsc{activate-new} appends a fresh frame
\((\ell,\emptyset,\emptyset)\) to \(K\). If such a frame already exists, the
rule \textsc{activate-existing} removes it from its current position and
re-appends it to the top of the stack. In both cases, the resulting top frame is
the frame labelled \(\ell\).

The remaining rules in \Cref{fig:frame-semantics} lift the byte-level semantics
to operations over a single stack frame. Each frame operation takes a frame
\(F=(\ell,B,\Sigma)\) and produces an updated frame while preserving the frame
label \(\ell\). The operation \(\operatorname{write}(F,o,w,bt)\) is defined when
the target interval \([o,o+w)\) lies inside the modeled byte array. It applies
the byte-array update \(B[o,w\leftarrow bt]\) and preserves the buffer metadata
\(\Sigma\).

The operation \(\operatorname{extend}(F,n)\) increases the size of the frame by
appending \(n\) fresh bytes in the \(\mathrm{Free}\) state. The operation
\(\operatorname{push}(F,n,bt)\) appends \(n\) initialized bytes to the frame:
\(\mathrm{RWrite}\) appends bytes in the \(\mathrm{Critical}\) state, whereas
\(\mathrm{nRWrite}\) appends bytes in the \(\mathrm{Occupied}\) state. The
operation \(\operatorname{pop}(F,n)\) removes the last \(n\) bytes from the
frame. The operation \(\operatorname{addbuf}(F,\sigma)\) records a stack-buffer
descriptor in the frame metadata without changing the byte array. Finally,
\(\operatorname{setcanary}(F)\) records the presence of a canary on the frame.

These frame operations define local updates to a single frame. Their effect on
the complete stack configuration is defined by the action semantics in
\Cref{fig:memstace-transition}. The semantic function

\[
    \llbracket a \rrbracket : \mathsf{Stack}\to\mathcal{P}(\mathsf{Stack})
\]

maps an abstract memory action \(a\) and a stack configuration \(K\) to the set
of possible successor stack configurations. For direct actions, the operation is
applied to the active frame \(F_n=\operatorname{top}(K)\), while the caller
frames \(F_0,\ldots,F_{n-1}\) are preserved. Therefore, \(\mathrm{Write}\),
\(\mathrm{Extend}\), \(\mathrm{Push}\), \(\mathrm{Pop}\), \(\mathrm{Canary}\)
and \(\mathrm{Buf}\) produce successor stacks of the form \(F_0\cdots F_n'\),
where \(F_n'\) is obtained by the corresponding frame operation in
\Cref{fig:frame-semantics}.

External calls and loops are handled through summary transitions.
\(\operatorname{ExtCallSummary}(c,\beta,K)\) returns the stack configurations
computed by the emulator for call \(c\) starting from stack \(K\), while
\(\operatorname{LoopSummary}(L,\beta,K)\) returns the stack configurations
computed for loop \(L\) starting from \(K\), with the symbolic execution bounds
\(\beta=(i,\tau,\eta)\), where $i$ bounds the number of loop iterations, $\tau$
bounds the number of symbolic execution steps, and $\eta$ bounds the number of
active symbolic states. Unlike direct frame operations, these summary
transitions are defined over complete stack configurations because the emulated
instructions may affect more than the active frame, we will discuss these
further in \Cref{ssec:emul}. The action \(\epsilon\) represents an instruction
with no relevant stack-memory effect and therefore leaves the stack unchanged.

Finally, the rule \textsc{Step} in \Cref{fig:memstace-transition} combines stack
updates with bCFG traversal. From a memory state \((\rho,K)\), the model checker
selects an outgoing bCFG edge \((\rho,\rho')\in E\), applies the abstract action
\(\alpha(\rho)\), and creates a successor state \((\rho',K')\) for every
\(K'\in\llbracket\alpha(\rho)\rrbracket(K)\). Thus, the program location evolves
according to the bCFG, while the stack configuration evolves according to the
semantics of the memory action at the current location.
\begin{figure}[t]
\centering
\small
\[
\textit{Stack operation.}
\qquad
\boxed{\mathit{Stack}_{\mathit{\,op}} : K \to K'}
\]
\begin{mathpar}
  \inferrule*[right=activate-new]
{ \ell \notin \operatorname{Labels}(K) } { \operatorname{activate}(K,\ell)
  \to
  K \cdot (\ell,\emptyset,\emptyset,\mathsf{false}) }

\inferrule*[right=activate-existing]
{ K = K_1 \cdot F \cdot K_2 \\
  F = (\ell,B,\Sigma,\gamma) \\
  \ell \in \operatorname{Labels}(K) } { \operatorname{activate}(K,\ell)
  \to
  K_1 \cdot K_2 \cdot F }
\end{mathpar}

\[
\textit{Frame operation.}
\qquad
\boxed{\mathit{Frame}_{\mathit{\,op}} : F \to F'}
\]
\begin{mathpar}

\inferrule*[right=write]
{ F = (\ell,B,\Sigma,\gamma) \\
  |B| = m \\
  0 \leq o \\
  o+w \leq m \\
  B' = B[o,w \leftarrow bt] } { \operatorname{write}(F,o,w,bt) \to
(\ell,B',\Sigma,\gamma) }

\inferrule*[right=extend]
{ n > 0 \\
  F = (\ell,B,\Sigma,\gamma) } { \operatorname{extend}(F,n)
  \to
  (\ell, B \cdot [\underbrace{\mathrm{Free},\ldots,\mathrm{Free}}_{n}], \Sigma,
\gamma) }

\inferrule*[right=push]
{ n > 0 \\
  F = (\ell,B,\Sigma,\gamma) \\
  \bigl(bt = \mathrm{RWrite} \land bs = \mathrm{Critical}\bigr)
  \lor
  \bigl(bt = \mathrm{nRWrite} \land bs = \mathrm{Occupied}\bigr) } {
\operatorname{push}(F,n,bt)
  \to
  (\ell, B \cdot [\underbrace{bs,\ldots,bs}_{n}], \Sigma, \gamma) }

\inferrule*[right=pop]
{ n > 0 \\
  F = (\ell,B \cdot C,\Sigma, \gamma) \\
  |C| = n } { \operatorname{pop}(F,n)
  \to
  (\ell,B,\Sigma, \gamma) }

\inferrule*[right=addbuf]
{ F = (\ell,B,\Sigma, \gamma) \\ \Sigma : \{\sigma_0, \dots, \sigma_n\} } {
\operatorname{addbuf}(F,\sigma)
  \to
  (\ell,B, \{\sigma_0,\dots,\sigma_n,\sigma\}, \gamma) }

\inferrule*[right=set-canary]
{ F = (\ell,B,\Sigma, \gamma) \\ \gamma = \mathit{false} } {
\operatorname{setcanary}(F)
  \to
  (\ell,B, \Sigma, \mathit{true}) }
\end{mathpar}
\vspace{1mm}
\hrule
\vspace{2mm}
\caption{Stack and Frame operation semantics.}
\label{fig:frame-semantics}
\end{figure}

\mypara{Memory State Space}
A \mss state records both a program location and an abstract stack:
\[
    s = (\rho,K).
\]
The memory state space of program $P$ is the transition system
\[
    \mss_P = (S,s_0,\longrightarrow),
\]
where
\[
    S \subseteq V \times \mathsf{Stack},
    \qquad
    s_0 = (\rho_0,K_0),
\]
and $K_0$ is the initial abstract stack. The transition relation
\(
\to \subseteq S\times S
\)
is defined by the \textsc{Step} rule in \Cref{fig:memstace-transition}: a
transition
\(
(\rho,K)\to(\rho',K')
\)
exists whenever
\(
(\rho,\rho')\in E
\)
and
\(
K' \in \llbracket \alpha(\rho)\rrbracket(K).
\)
\begin{figure*}[t]
\centering
\small
\setlength{\arraycolsep}{4pt}
\renewcommand{\arraystretch}{1.15}

\[
\boxed{\llbracket a \rrbracket : \mathsf{Stack} \to \mathcal{P}(\mathsf{Stack})}
\qquad
\frac{
  (\rho,\rho') \in E
  \qquad
  K' \in \llbracket \alpha(\rho) \rrbracket(K)
}{
  (\rho,K) \longrightarrow (\rho',K')
}
\;\textsc{Step}
\]

\[
\begin{array}{@{}l@{\qquad}l@{}}
\begin{aligned}[t]
\llbracket \mathsf{Write}(o,w,bt) \rrbracket(F_0 \cdots F_n)
\triangleq{}&
\{\,F_0 \cdots F_n'
  \mid \operatorname{write}(F_n,o,w,bt) \to F_n'\,\}
\\[0.6em]
\llbracket \mathsf{Extend}(w) \rrbracket(F_0 \cdots F_n)
\triangleq{}&
\{\,F_0 \cdots F_n'
  \mid \operatorname{extend}(F_n,w) \to F_n'\,\}
\\[0.6em]
\llbracket \mathsf{Push}(w,bt) \rrbracket(F_0 \cdots F_n)
\triangleq{}&
\{\,F_0 \cdots F_n'
  \mid \operatorname{push}(F_n,w,bt) \to F_n'\,\}
\\[0.6em]
\llbracket \mathsf{Pop}(w) \rrbracket(F_0 \cdots F_n)
\triangleq{}&
\{\,F_0 \cdots F_n'
  \mid \operatorname{pop}(F_n,w) \to F_n'\,\}
\\[0.6em]
\llbracket \mathsf{Buf}(\sigma) \rrbracket(F_0 \cdots F_n)
\triangleq{}&
\{\,F_0 \cdots F_n'
  \mid \operatorname{addbuf}(F_n,\sigma) \to F_n'\,\}
\end{aligned}
&
\begin{aligned}[t]
\llbracket \mathsf{ActivateFrame}(\ell) \rrbracket(K)
\triangleq{}&
\{\,K' \mid \operatorname{activate}(K,\ell) \to K'\,\}
\\[0.6em]
\llbracket \mathsf{Canary} \rrbracket(F_0 \cdots F_n)
\triangleq{}&
\{\,F_0 \cdots F_n'
  \mid \operatorname{setcanary}(F_n) \to F_n'\,\}
\\[0.6em]
\llbracket \mathsf{ExtCall}(c,\beta) \rrbracket(K)
\triangleq{}&
\{\,K' \mid (K,K') \in \operatorname{ExtCallSummary}(c,\beta,K)\,\}
\\[0.6em]
\llbracket \mathsf{Loop}(L,\beta) \rrbracket(K)
\triangleq{}&
\{\,K' \mid (K,K') \in \operatorname{LoopSummary}(L,\beta,K)\,\}
\\[0.6em]
\llbracket \epsilon \rrbracket(K)
\triangleq{}&
\{\,K\,\}.
\end{aligned}
\end{array}
\]
\hrule
\vspace{2mm}
\caption{Action semantics and MemStaCe transition relation.}
\label{fig:memstace-transition}
\end{figure*}

\subsection{Summary Transitions for External Calls and Loops}
\label{ssec:emul}

The direct transition rules described above model stack effects that can be
derived from a single assembly instruction, such as stack writes, pushes, pops,
and frame extensions. However, not all stack effects are visible from a single
instruction. Calls to external functions and loop bodies may modify the stack
through a sequence of instructions whose effects depend on calling-convention
arguments, path conditions, and execution bounds. \tool represents these through
\emph{summary transitions}.

A summary transition is a relation over complete stack configurations. Given a
pre-summary stack configuration \(K\), an external-call summary computes the
possible successor stack configurations after the summarized call, while a loop
summary computes the possible successor stack configurations after a bounded
number of loop iterations. These summaries are defined as
\begin{mathpar}
    \operatorname{ExtCallSummary}(c,\beta,K) \subseteq \mathsf{Stack} \\
    \operatorname{LoopSummary}(L,\beta,K) \subseteq \mathsf{Stack},
\end{mathpar}
where \(c=\rho\) identifies the program location of the external call site in
the recovered bCFG, and \(L=\rho\) identifies the program location of the loop
header or entry point being summarized. The parameter \(\beta=(i,\tau,\eta)\)
defines the symbolic-execution bounds used during summary construction, where
\(i\) bounds the number of loop iterations, \(\tau\) bounds the number of
symbolic execution steps used by angr, and \(\eta\) bounds the number of active
symbolic states maintained during the analysis. Although the iteration bound
\(i\) is only relevant to loop summaries, \tool uses the same parameterization
\(\beta=(i,\tau,\eta)\) for both call and loop summaries to maintain a uniform
summary interface.
The stack configuration $K$ is the current abstract stack at the program point
where a call or loop summary is applied. Therefore, summaries are local to the
current state-space construction path; they are not computed from a global
initial stack.

\mypara{User functions}
\tool distinguishes user-functions whose code is present in the target binary
from external library calls. These functions are not treated as black boxes,
since they appear as assembly code in the recovered bCFG. When the model checker
reaches a call to a function whose implementation is present in the recovered
bCFG, \tool creates or reuses the corresponding abstract stack frame and
continues traversing the callee's assembly instructions. The stack effects of
such functions are then derived by applying the same \mss transition rules used
for the rest of the binary.

\mypara{External C library calls}
For calls whose implementation is not traversed as part of the target binary,
\tool uses external-call summaries. When the model checker encounters a call
instruction, \tool first determines whether the target is a recovered
user-function or an external function. For the latter, \tool identifies the
function name from the bCFG and consults its C library function database to
recover the expected argument types. It then performs a local reverse-flow
analysis within the basic block containing the call instruction to recover the
values assigned to the System V argument registers, such as \texttt{RDI},
\texttt{RSI}, \texttt{RDX}, \texttt{RCX}, \texttt{R8}, and \texttt{R9}. For
instance, in a call to \texttt{strcpy}, the destination pointer is expected in
\texttt{RDI} and the source pointer in \texttt{RSI}. If a recovered argument
points to a buffer in the current stack frame, \tool records the corresponding
buffer offset and size.
\tool first attempts to summarise supported unsafe C library calls statically.
If the destination argument and the relevant stack-buffer metadata can be
recovered, the model analysis computes the interval of stack bytes that may be
written by the call. This interval is then propagated to the active frame using
the byte transition semantics defined earlier.
For example, a recovered destination buffer for \texttt{strcpy} allows \tool to
compute which bytes of the current frame may be affected by the copy. For
\text{memcpy} and \texttt{memmove}, the length argument is used when it can be
recovered; otherwise, the summary must fall back to symbolic execution.

\mypara{Bounded Symbolic Execution}
If static recovery is insufficient, \tool falls back to the bounded angr-based
symbolic execution. The bounded executer searches for a program state that
reaches the target call site from the selected analysis entry. This search is
bounded by the step limit \(\tau\) and active-state limit \(\eta\). When a state
reaches the target, \tool records the relevant stack bytes immediately before
the call, executes the call, records the stack bytes again, and compares the
contents of both stacks. The byte positions that differ are converted into \mss
by the updates and applied to the current abstract stack frame.
These states that reach their respective target are cached by the analysis,
which records both start and target addresses. This avoids repeating the same
exploration whenever multiple summaries require the same target state. However,
this execution-based summary is still bounded. If the target address cannot be
reached within the configured limits, if symbolic execution produces only
unusable states, or if the post-call stack cannot be inspected, the summary is
marked as failed.

\mypara{Concolic Inputs}
When a summarised function consumes data from \texttt{stdin} or \texttt{argv},
\tool records an input associated with the summary. In the bounded symbolic
execution path, the input is obtained by concretising symbolic bytes: \tool
queries the solver for a satisfying assignment under the path constraints that
lead to the summarised call or stack write, and saves the resulting concrete
byte string. When the static-summary path is used, no symbolic execution state
is available, so \tool synthesises a stress-test input instead. These inputs are
later used during patch validation as a candidate malicious, boundary, or
stress-test inputs, but they are not claimed to be complete exploit witnesses
for all executions.

\mypara{Loop summaries}
Loops are also represented through bounded summary transitions. \tool first
identifies loop headers and exits from the recovered bCFG. For a loop \(L\), the
bounded symbolic executor attempts to reach the loop entry. Once the loop entry
is reached, \tool records the stack contents, executes the loop until either a
loop exit is reached or the iteration bound \(i\) is exhausted, and then records
the stack contents again. The differences between the pre-loop and post-loop
contents are translated into byte updates in the corresponding \mss frame.
If a loop requires more than \(i\) iterations to expose a buffer overflow, or if
the loop entry or exit cannot be reached within the configured angr bounds,
\tool may fail to compute a complete loop summary. In that case, the analysis
continues with the current abstract stack state \(K\), but does not claim
completeness for stack effects hidden behind the failed loop summary.

\section{Model Checking and Vulnerability Detection}
\label{sec:secProper}
This section describes how \tool checks stack-security properties over the
\mss constructed in \Cref{sec:model_checker}. First, it defines the query
functions used to observe \mss states and explains how these observations give
rise to atomic propositions for LTL. It then presents the predefined LTL
properties used to capture stack-buffer-overflow patterns. Finally, it describes
how \tool checks these properties using Büchi automata and how violations are
reported as counterexample traces that identify the program locations and
memory actions responsible for the unsafe write.

\subsection{From MemStaCe States to Atomic Propositions}
\label{ssec:memstace_predicates}
\tool checks LTL properties over atomic propositions interpreted on \mss
states. The query functions in \Cref{fig:memstace_query_functions} are not
atomic propositions themselves; instead, they provide the state-dependent
observations from which atomic propositions are defined. For a memory state
\(s=(\rho,K)\), an atomic proposition is a Boolean predicate over the current
program location, the current stack configuration, and the transition by which
\(s\) was reached.
Although a state is written as \(s=(\rho,K)\), \tool treats \mss states as
nodes of the unfolded exploration graph. Thus, two distinct nodes may carry the
same pair \((\rho,K)\) if they are reached through different paths. This
convention makes \(\mathsf{PreviousTrans}\) well-defined for every non-initial
state.
When specifications quantify over \mss objects, the quantification is bounded by
the finite sets returned by these query functions. For example, a formula of the
form
\[
    \forall F \in \mathsf{Frames}(K).\ \varphi(F)
\]
is interpreted at the current state as a finite conjunction over the frames
currently represented in \(K\). Therefore, the underlying temporal reasoning
remains finite-state, while the specification language can refer directly to
frames, buffers, and byte offsets.
\Cref{fig:memstace_query_functions} defines the query functions used by
\tool. The function \(\mathsf{Frames}(K)\) returns the set of abstract stack
frames present in the current state, and \(\mathsf{Labels}(K)\) returns their
labels. Assuming labels are unique within \(K\), \(\mathsf{Stack}(K,\ell)\)
returns the frame associated with label \(\ell\). For a frame
\(F=(\ell,B,\Sigma,\gamma)\), the function \(\mathsf{Byte}(F,i)\) returns the
abstract byte state stored at offset \(i\), while \(\mathsf{HasCanary}(F)\)
returns the canary metadata \(\gamma\).

\begin{figure}[!b]
\centering
\small
\[
\begin{array}{rcl}
\operatorname{Frames}(K)
  &\triangleq&
  \{F \mid F=(\ell,B,\Sigma, \gamma) \text{ is present in } K\}
\\[0.4em]

\operatorname{Labels}(K)
  &\triangleq&
  \{\ell \mid \exists B,\Sigma,\gamma.\;(\ell,B,\Sigma,\gamma)\in\mathsf{Frames}(K)\}
\\[0.4em]

\operatorname{Stack}(K,\ell)
  &\triangleq&
  F
  \quad\text{where } F=(\ell,B,\Sigma,\gamma)\in\mathsf{Frames}(K)
\\[0.4em]

\operatorname{Byte}(F,i)
  &\triangleq&
  B(i)
  \quad\text{where } F=(\ell,B,\Sigma,\gamma)
\\[0.4em]

\operatorname{HasCanary}(F)
  &\triangleq&
  \gamma
  \quad\text{where } F=(\ell,B,\Sigma,\gamma)
\\[0.4em]

\operatorname{Buffers}(F)
  &\triangleq&
  \Sigma
  \quad\text{where } F=(\ell,B,\Sigma,\gamma)
\\[0.4em]

\operatorname{BuffStart}(\sigma)
  &\triangleq& o
  \quad\text{where } \sigma=(o,n)
\\[0.4em]

\operatorname{BuffEnd}(\sigma)
  &\triangleq&
  o+n-1
  \quad\text{where } \sigma=(o,n)
\\[0.4em]

\operatorname{BuffSize}(\sigma)
  &\triangleq&
  n
  \quad\text{where } \sigma=(o,n)
\\[0.4em]

\operatorname{PreviousTrans}(s)
  &\triangleq&
  \begin{cases}
    \bot, & \text{if } s=s_0,\\
    \alpha(\rho_p), &
    \text{where }
    \exists!\,(\rho_p,K_p).\;
    (\rho_p,K_p)\longrightarrow s.
  \end{cases}
\end{array}
\]
\hrule
\vspace{2mm}
\caption{MemStaCe query functions.}
\label{fig:memstace_query_functions}
\end{figure}

The function \(\mathsf{Buffers}(F)\) returns the buffer metadata associated with
frame \(F\). Accordingly, \(\mathsf{BuffStart}\), \(\mathsf{BuffEnd}\), and
\(\mathsf{BuffSize}\) expose the start offset, inclusive end offset, and size of
a buffer descriptor. These functions allow formulas to relate byte states to
buffer boundaries.
Finally, \(\mathsf{PreviousTrans}(s)\) returns the abstract memory action by
which the current state was reached. This query is useful for properties that
depend on the last semantic step, for example, whether the current state was
reached by an external-call summary or by a loop summary.

\begin{figure*}[!t]
\centering
\small

\begingroup
\setlength{\jot}{2pt}
\setlength{\abovedisplayskip}{3pt}
\setlength{\belowdisplayskip}{3pt}

\def\propindent{1.05em}
\def\halfcol{0.475\textwidth}

\long\def\Prop#1#2{%
  \par\vspace{2pt}%
  {\footnotesize\scshape #1}\par\vspace{1pt}%
  \noindent\hspace*{\propindent}%
  $\begin{aligned}[t]#2\end{aligned}$%
  \par\vspace{4pt}%
}

\long\def\Panel#1#2{%
  \begin{tcolorbox}[
    enhanced,
    width=\linewidth,
    colback=white,
    colframe=black,
    colbacktitle=white,
    coltitle=black,
    boxrule=-1pt,
    titlerule=-1pt,
    arc=0pt,
    outer arc=0pt,
    boxsep=0pt,
    left=7pt,
    right=7pt,
    top=6pt,
    bottom=6pt,
    toptitle=3pt,
    bottomtitle=3pt,
    lefttitle=7pt,
    righttitle=7pt,
    before skip=0pt,
    after skip=7pt,
    fonttitle=\bfseries\scshape\footnotesize,
    title={#1},
    before upper={\parindent=0pt},
  ]
  #2
  \end{tcolorbox}%
}

\noindent
\begin{minipage}[t]{\halfcol}
\vspace{0pt}

\Panel{Control-Data Integrity}{%
\Prop{RIP Integrity}{
  &\Box\Bigl(
    \forall F\in\mathsf{Frames}(K).\;
    \textstyle\bigwedge_{i=0}^{7}
    \mathsf{Byte}(F,i)=\mathsf{C}
  \Bigr)
}

\Prop{RBP Integrity}{
  &\Box\Bigl(
    \forall F\in\mathsf{Frames}(K).\;
    \textstyle\bigwedge_{i=8}^{15}
    \mathsf{Byte}(F,i)=\mathsf{C}
  \Bigr)
}

\Prop{Canary Integrity}{
  &\Box\Bigl(
    \forall F\in\mathsf{Frames}(K).\;
    \mathsf{HasCanary}(F)\Rightarrow{}\\
  &\quad
    \textstyle\bigwedge_{i=16}^{23}
    \mathsf{Byte}(F,i)=\mathsf{C}
  \Bigr)
}

\Prop{No Off-by-One Overflow}{
  &\Box\neg\Bigl(
    \exists F\in\mathsf{Frames}(K).\\
  &\quad
    \mathsf{Byte}(F,15)=\mathsf{M}
    \land
    \mathsf{Byte}(F,14)=\mathsf{C}
  \Bigr)
}
\vspace{1em}
}

\Panel{Unsafe Function Violations}{%

\Prop{No \texttt{gets} Usage}{
  &\Box\neg\,\mathsf{call}(\mathtt{gets})
}

}

\end{minipage}%
\hfill
\begin{minipage}[t]{\halfcol}
\vspace{0pt}

\Panel{Stack-Buffer-Boundary Violations}{%

\Prop{No Stack Underwrite}{
  &\Box\neg\Bigl(
    \exists F\in\mathsf{Frames}(K).\;
    \exists\sigma\in\mathsf{Buffers}(F).\\
  &\quad
    \mathsf{BuffSize}(\sigma)>31
    \land{}\\
  &\quad
    \mathsf{Byte}
      (F,\mathsf{BuffStart}(\sigma){+}1)
    =
    \mathsf{M}
    \land{}\\
  &\quad
    \mathsf{Byte}
      (F,\mathsf{BuffStart}(\sigma){+}2)
    =
    \mathsf{O}
  \Bigr)
}

\Prop{No Buffer Overflow by One}{
  &\Box\Bigl(
    \mathsf{summary}
    \Rightarrow{}\\
  &\quad
    \neg\Bigl(
      \exists F\in\mathsf{Frames}(K).\;
      \exists\sigma\in\mathsf{Buffers}(F).\\
  &\qquad
      \mathsf{Byte}
        (F,\mathsf{BuffEnd}(\sigma))
      =
      \mathsf{O}
      \land{}\\
  &\qquad
      \mathsf{Byte}
        (F,\mathsf{BuffEnd}(\sigma){-}1)
      =
      \mathsf{M}
    \Bigr)
  \Bigr)
}

\Prop{No Underflow}{
  &\Box\neg\Bigl(
    \mathsf{summary}
    \;\mathcal{R}\;
    \Diamond\Bigl(\\
  &\quad
    \exists F\in\mathsf{Frames}(K).\;
    \exists\sigma\in\mathsf{Buffers}(F).\\
  &\quad
    \mathsf{Byte}
      (F,\mathsf{BuffStart}(\sigma))
    =
    \mathsf{O}
    \land{}\\
  &\quad
    \mathsf{Byte}
      (F,\mathsf{BuffStart}(\sigma){+}1)
    =
    \mathsf{O}
    \land{}\\
  &\quad
    \mathsf{Byte}
      (F,\mathsf{BuffStart}(\sigma){+}2)
    \neq
    \mathsf{O}
    \Bigr)
  \Bigr)
}

}

\end{minipage}

\vspace{1pt}

\noindent
\begin{minipage}[t]{\textwidth}
\vspace{2pt}

\Panel{Auxiliary Definitions and Abbreviations}{%

\centering
$\displaystyle
\begin{gathered}
\mathsf{C}\equiv\mathrm{Critical},
\quad
\mathsf{M}\equiv\mathrm{Modified},
\quad
\mathsf{O}\equiv\mathrm{Occupied},
\\[4pt]
\begin{aligned}
\mathsf{call}(f)
&\triangleq
\exists c,\beta.\;
\mathsf{PreviousTrans}(s)=\mathsf{ExtCall}(c,\beta)
\land
\mathsf{inst}(c)=\mathtt{call}\; f,
\\[3pt]
\mathsf{summary}
  &\triangleq
  \bigl(
    \exists c,\beta.\;
    \mathsf{PrevTrans}(s)=\mathsf{ExtCall}(c,\beta)
  \bigr)
  \lor
  \bigl(
    \exists L,\beta.\;
    \mathsf{PrevTrans}(s)=\mathsf{Loop}(L,\beta)
  \bigr).
\end{aligned}
\end{gathered}
$

}

\end{minipage}

\vspace{1mm}
\hrule
\vspace{2mm}
\caption{\tool LTL security properties. Auxiliary definitions and
abbreviations used by the formulas are given in the bottom panel.}
\label{fig:basics_ltl_properties}

\endgroup
\end{figure*}

\subsection{Security Properties}
\label{sec:sec_properties}

Since BASICS targets stack-based buffer overflows, we define eight LTL security
properties, shown in \Cref{fig:basics_ltl_properties}. These properties are
divided into three classes: 
\emph{Control-Data Integrity} protect 
control-sensitive stack data, namely the saved return address (\texttt{RIP}), saved
base pointer (\texttt{RBP}), and stack canary, when present. They require the
corresponding byte ranges in each stack frame to remain \(\mathsf{C}\), i.e.,
critical. \emph{Unsafe Function Violations} captures calls to intrinsically
unsafe library functions, currently \texttt{gets}. \emph{Stack-Buffer-Boundary Violations} reason about writes around recovered stack-buffer boundaries, including off-by-one overwrites, underwrites, overflow-by-one patterns, and underflows caused by summarised calls or loops. Next, we give an overview of the security properties they comprise. 

\mypara{Control-Data Integrity}
The \emph{RIP Integrity} property checks that the saved return-address region of
each stack frame remains critical throughout the execution. A violation
indicates that a write has reached control-flow data and may corrupt the return
target of the current function. The \emph{RBP Integrity} property is analogous,
but protects the saved base pointer. Although overwriting \texttt{RBP} does not
always immediately redirect control flow, it corrupts the caller's frame layout
and is a common symptom of stack-frame overflow. The \emph{Canary Integrity} property applies only to frames in which BASICS recovers a stack canary; it
requires the canary byte range to remain critical, and therefore flags writes
that would bypass or corrupt compiler-inserted stack protection.
The \emph{No Off-by-One Overflow} property captures a narrow control-data
corruption pattern in which a write modifies a byte adjacent to critical stack
metadata. This property is useful for detecting small overflows that may not
overwrite the full return-address or base-pointer range, but still cross the
boundary into protected stack data.

\mypara{Unsafe Function Violations}
The \emph{No Gets Usage} property is a separate unsafe-function property: because \texttt{gets} performs unbounded
input into a destination buffer, BASICS flags any transition corresponding to a
call to \texttt{gets}.

\mypara{Stack-Buffer-Boundary Violations}
The remaining properties are related to the boundaries of the recovered stack
buffers. \emph{No Stack Underwrite} detects writes near the beginning of a
buffer that disturb the expected byte-state pattern at the lower boundary. The
\emph{No Buffer Overflow by One} property checks whether a loop or call summary
has modified bytes immediately adjacent to the end of a recovered buffer,
capturing one-byte overwrite patterns. Finally, \emph{No Underflow} detects
summary-induced writes that move before the start of a recovered buffer.
\color{black}

\subsection{Detecting Vulnerabilities}

\label{ssec:detecting_vulns}

\mypara{Property Checking}
The transition system \(\mss_P=(S,s_0,\to)\) defines the memory states explored
by the model checker. To check an LTL property \(\varphi\), \tool derives a
property-specific Kripke structure
\[
    \mathcal K_{P,\varphi}
    =
    (S,s_0,R,AP_\varphi,\mathcal L_\varphi)
\]
from \(\mss_P\). This construction keeps the same states and initial state,
adds stuttering self-loops to terminal states, and labels each state with the
atomic propositions of \(\varphi\) that hold there.
The set \(AP_\varphi\) contains the atomic propositions that occur in
\(\varphi\). Each proposition is a Boolean predicate over \mss states, evaluated using the query functions in \Cref{fig:memstace_query_functions}. Thus,
\[
    \mathcal L_\varphi(s)
    =
    \{\,p\in AP_\varphi \mid \llbracket p \rrbracket_s = \mathsf{true}\,\}.
\]
The query functions provide the observation interface over \mss states, while
\(AP_\varphi\) contains the Boolean propositions used by the LTL checker.
The Kripke transition relation \(R\) is the totalised version of the \mss
transition relation:
\[
    (s,s')\in R
    \quad\text{iff}\quad
    s\longrightarrow s'
    \;\lor\;
    \bigl(s=s' \land \nexists t\in S.\; s\longrightarrow t\bigr).
\]
The second case adds a stuttering self-loop to terminal states, allowing finite
\mss executions to be interpreted as infinite paths, as required by LTL, without
adding any memory effect.
Each property \(\varphi\) is translated into a Büchi automaton, and \tool checks
\(\mathcal K_{P,\varphi}\) against the automaton for \(\neg\varphi\). A
reachable accepting run corresponds to an execution of the program abstraction
that violates \(\varphi\). Since the predefined properties are safety-oriented invariants, violations are witnessed by finite bad prefixes.
In the implementation, \tool does not explicitly construct the full product
graph. Instead, it traverses \(\mss_P\) while maintaining the current automaton state for each property. At each memory state, the automaton guards are
evaluated through \(\mathcal L_\varphi\), and the automaton state is advanced
accordingly.

\mypara{Counterexample Traces}
A vulnerability candidate is reported when the automaton for a security property
reaches a violating configuration along a reachable \mss path. The report
contains the violated property and a counterexample trace that explains how the
violating state was reached.
A trace is a sequence of tuples
\[
    \langle \mathit{addr}, \mathit{instr}, \mathit{MemAct} \rangle,
\]
where \(\mathit{addr}\) is the instruction address, \(\mathit{instr}\) is the
disassembled instruction, and \(\mathit{MemAct}\) is the static memory action
or summary action applied at that step. The trace links the LTL violation back
to concrete binary locations and to the stack-memory updates that caused the
violation.

\begin{figure}[t]
    \centering
    \includegraphics[width=\columnwidth]{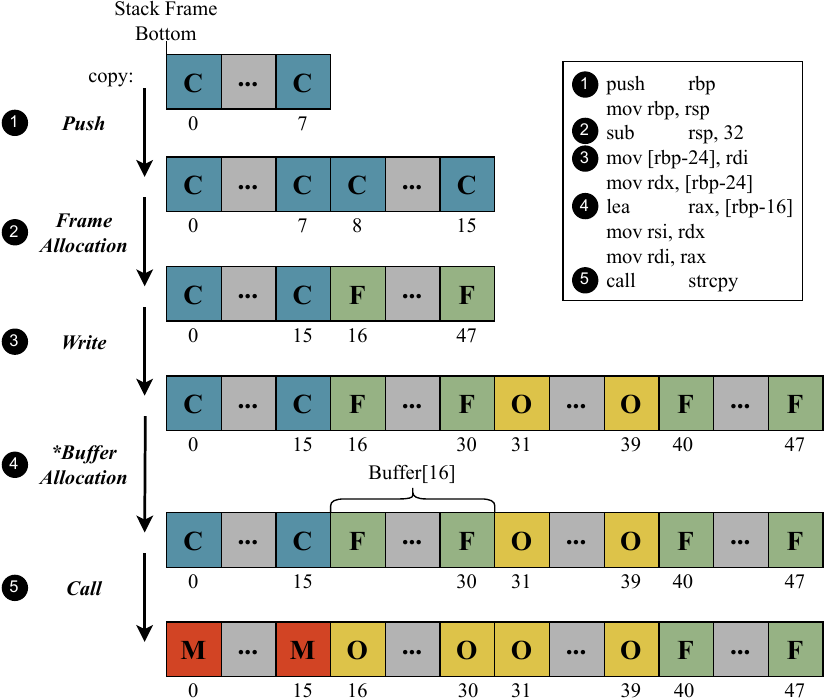}
    \caption{Example of the \mss for the \texttt{copy} function in Listing
    \ref{copy_function_asm}, where \textbf{C}: Critical, \textbf{O}: Occupied,
    \textbf{F}: Free, \textbf{M}: Modified.}
    \label{fig:mss}
\end{figure}

\mypara{Example} 
To illustrate the creation of the \mss and the detection of a security property
violation, Figure~\ref{fig:mss} shows the step-by-step construction of the \mss
for the \texttt{copy} function (Listing~\ref{copy_function_asm}). The process
begins by initialising the stack frame with 8 bytes in the \textit{Critical}
state, and then pushing the stack base pointer onto the stack frame
\blackbubble{1}. The frame is then extended \blackbubble{2}; newly allocated
bytes are marked \textit{Free}, reflecting the uninitialized space. Later,
instruction \blackbubble{3} prepares the call to \texttt{strcpy}, and although instruction \blackbubble{4} does not modify memory directly, it represents a transition that identifies a target buffer address for later use. The final instruction \blackbubble{5} performs the actual call to \texttt{strcpy}, whose effect is summarised in the \mss. However, due to a buffer overflow, the write extends into adjacent memory and modifies \textit{Critical} bytes, transitioning them to the \textit{Modified} state. This last transition triggers a violation of the RIP Integrity Property, resulting in the trace being flagged by the model checker as a counterexample.
Listing~\ref{trace} shows the counterexample trace for the same violation. The trace contains the frame setup, the local stack allocation, and the call to \texttt{strcpy}. The call summary writes beyond the recovered destination buffer and modifies bytes in the saved return-address range. Since \emph{RIP Integrity} requires these bytes to remain critical in all reachable states, \tool reports the trace as a counterexample.
\begin{listing}[t]
\begingroup\fvset{highlightcolor=gray!20,frame=lines, fontsize=\footnotesize, framesep=2mm, breaklines, highlightlines={7}}\begin{MintedVerbatim}[commandchars=\\\{\}]
\PYGZlt{}0x401169; endbr64; ActivateFrame(\PYGZdq{}copy\PYGZdq{})\PYGZgt{}
\PYGZlt{}0x40116d; push rbp; Push(8, RWrite)\PYGZgt{}
\PYGZlt{}0x401171; sub rsp, 0x20; Extend(32)\PYGZgt{}
\PYGZlt{}0x401175; mov dword ptr [rbp\PYGZhy{}0x14], edi; Write(19, 4, nRWrite)\PYGZgt{}
\PYGZlt{}0x401178; mov qword ptr [rbp\PYGZhy{}0x20], rsi; Write(31, 8, nRWrite)\PYGZgt{}
\PYGZlt{}0x401187; lea rax, [rbp\PYGZhy{}0xa]; Buf((10, 2))\PYGZgt{}
\PYGZlt{}0x401191; call 0x401060 \PYGZlt{}strcpy\PYGZgt{}; ExtCall(0x401191)\PYGZgt{}
\end{MintedVerbatim}
\endgroup\caption{Counter-Example Trace
   for a program that violates the \emph{RIP Integrity} Security Property.}
\label{trace}
\vspace{-5mm}
\end{listing}

\subsection{\DIFF{Bounded Soundness and Completeness Conditions}}
\label{ssec:bounded_correctness}

The model-checking result is relative to the constructed \mss. \tool therefore does not claim unconditional soundness or completeness for arbitrary binary executions. \DIFF{Instead, the following conditions characterise the validity of counterexamples relative to the constructed abstraction and the circumstances under which concrete stack-buffer overflows are represented by the bounded analysis.}

\newtheorem{theorem}{Theorem}
\newtheorem{condition}{Condition}

\begin{condition}[\DIFF{Bounded soundness condition}]
Let \(\pi\) be a counterexample trace reported by \tool for property
\(\varphi\). If the recovered bCFG contains the corresponding program path, the
instruction-to-action translation is correct for every instruction in \(\pi\),
and all call and loop summaries used in \(\pi\) accurately represent the stack
bytes written by the corresponding bounded execution segments, then \(\pi\)
witnesses a violation of \(\varphi\) in the bounded \mss abstraction.
\end{condition}

\paragraph{\DIFF{Argument}}
By construction, each transition in \(\pi\) follows an edge of the recovered
bCFG and applies the action semantics defined in
Section~\ref{sec:model_checker}. Under the stated assumptions, these actions
preserve the relevant stack-byte effects of the corresponding binary
instructions or summaries. The labelling function exposes these byte states and
state predicates to the property automaton. Therefore, if the automaton reaches
a violating configuration along \(\pi\), the corresponding sequence of \mss
states violates \(\varphi\).

\begin{condition}[\DIFF{Bounded completeness condition}]
Assume a concrete stack-buffer overflow occurs along an execution path contained
in the recovered bCFG, within the configured loop bound and angr execution
limits, and through instructions or summarised calls whose stack effects are
captured by \tool. If the overflow changes byte states in a way covered by one
of the predefined security properties, then the constructed \mss contains a
corresponding violating path, \DIFF{which \tool can report as a counterexample.}
\end{condition}

\paragraph{\DIFF{Argument}}
The assumptions ensure that the relevant execution path is represented in the
recovered bCFG and that each stack effect along the path is reflected by a direct \mss transition or a successful summary transition. Therefore, the
write that causes the overflow appears as a modified byte pattern in some
reachable \mss state. Since the predefined property is evaluated on these
byte states, the corresponding automaton detects the violation, and the traversal
can emit the associated counterexample trace.

These conditions are bounded by design. \tool may miss vulnerabilities whose
paths are absent from the recovered bCFG, whose effects occur beyond the loop
iteration bound, whose calls cannot be summarised, or whose stack writes are
hidden by unsupported instructions, unresolved indirect calls, or failed
argument recovery. Conversely, conservative summaries may introduce false
positives when a statically possible overwrite is infeasible in the concrete
program. We empirically evaluate these cases in Section~\ref{sec:evaluation}.

\section{Patching Vulnerabilities and Bounded Validation}
\label{sec:patcher}

When the model checker reports a security-property violation, \tool attempts to repair the binary if the counterexample can be linked to a supported patchable sink. 
\DIFF{In the case of multiple property violations identifying the same
call site, \tool consolidates them into a single patch sink. On the other hand, distinct supported call sites are processed independently, and the patch template is selected according to the normalised called-function name.} %
The repair pipeline has four stages: identifying the sink responsible for the violation, generating a patch payload for that sink, installing the payload in the binary, and validating the resulting patched binary. 
The validation step is bounded: it provides evidence that the patch removes the observed vulnerable behaviour and preserves behaviour on tested benign and boundary inputs, but it does not prove whole-program functional equivalence.

\mypara{Vulnerability Identification} 
Using the counterexample trace produced for the violated security property, a reverse flow analysis is performed, tracing the execution backwards to identify the instruction behind the violation, as well as its location in bCFG and the user function address that contains such instruction. Since most BOs originate from function calls, the vulnerability is often linked to the \texttt{call} instruction at the end of the trace. For example, in the counterexample of Listing~\ref{trace}, this process starts from the end of the counterexample, following a backwards analysis until finding the instruction \texttt{call 0x401060}, a function call presented in BO exploitation. For vulnerabilities that are not related to external function calls, such as buffer underflows caused by loops, the same reverse analysis is applied, but looking for the entry point of the loop instead of a function call. 
Also, using the \textit{User Function Map}, the name of the user function is obtained by searching for the user function address extracted from the backwards analysis. 
Additionally, \DIFF{to support developers and analysts with the types of BOs identified,} the Common Weakness Enumeration (CWE) classes are provided for the properties that are violated. For this, the \textit{Vulnerability CWE Class} database, presented in \Cref{tab:cwe_map}, contains a map from security properties to \DIFF{one or more CWE classes}. 
\DIFF{For example, the return address integrity property (RIP Integrity in
\Cref{fig:basics_ltl_properties}) is associated with CWE-121 and
CWE-787~\citep{cwe121,cwe787}, representing stack-based buffer overflow and
its more general out-of-bounds-write class, respectively.}

\begin{table}[b]\footnotesize
\caption{\small Mapping of Security Properties to CWE Classes.}
\vspace{-3mm}
\centering
 \begin{tabular}{@{}ll@{}}
  \toprule
  \textbf{Security Property} & \textbf{CWE Class} \\ \midrule
  RIP Integrity                  & CWE-121, CWE-787 \\
  RBP Integrity                  & CWE-121, CWE-787 \\
  Canary Integrity               & CWE-121, CWE-787 \\
  No \texttt{gets} Usage         & CWE-676 \\
  No Off-by-One Overflow         & CWE-193, CWE-121, CWE-787 \\
  No Buffer Overflow by One      & CWE-193, CWE-119, CWE-121, CWE-787 \\
  No Stack Underwrite            & CWE-124 \\
  No Underflow                   & CWE-124 \\
  \bottomrule
  \end{tabular}
\label{tab:cwe_map}
\end{table}

\mypara{Patch Template Generation}
For each patchable sink, \tool selects a patch template according to the
recovered library function. 
The template is instantiated using the pre-call state
recovered during MemStaCe construction and call-summary generation. This state
contains the relevant System~V AMD64 argument registers, the destination object,
the recovered destination size when available, stack-canary metadata, and the
continuation address after the replaced call.

Templates have two variants. If the destination stack-buffer size is recovered
statically, the known-size variant receives this bound directly. If the bound
cannot be recovered statically, the unknown-size variant computes a conservative
runtime bound from the stack-frame layout, using \texttt{RBP}, the destination
pointer, and the canary offset when a canary is present. Runtime-bound templates
are used only when the destination is identified as a stack object in the
current frame; \tool skips destinations known to be heap-backed or otherwise
outside the supported stack-object model. Patch payloads also preserve the
expected library-call return convention, so that callers observing the return
value see the payload as a replacement for the original call.

All templates follow the same replacement pattern. The inserted payload is
called instead of the unsafe library routine, receives the original call
arguments, and enforces a bound before writing to the destination object. String
templates limit the copy to the destination capacity and explicitly write a
trailing \texttt{NUL}; concatenation templates also account for bytes already
present in the destination; raw-copy templates clamp the requested length; and
input templates bind the token or formatted assignment that writes into the
stack object. After the payload finishes, execution resumes at the instruction
following the original call. Thus, the patch changes the unsafe write behaviour
without changing the surrounding control flow.

Listing~\ref{l:strcpy_patch} shows the statically known-size \texttt{strcpy} payload. The
payload bounds the copy to \texttt{size - 1}, forces a terminating NUL byte,
handles degenerate bounds, and writes the expected \texttt{strcpy} return value
back to the caller-visible return slot. When the destination buffer size cannot be recovered
statically, the patch approximates a runtime bound by computing the distance, in bytes, between
\texttt{RBP} and \texttt{RDI}. Listing~\ref{l:strcpy_patch_dynamic} shows this dynamic variant.

\begin{listing}[t]
   \begingroup\fvset{highlightcolor=gray!20,frame=lines, numbers=left, numbersep=-3pt, xleftmargin=0pt, breaklines, breakanywhere, fontsize=\footnotesize, framesep=2mm}\begin{MintedVerbatim}[commandchars=\\\{\}]
\PYG{+w}{  }\PYG{c+cp}{\PYGZsh{}include}\PYG{+w}{ }\PYG{c+cpf}{\PYGZdq{}stdlib.c\PYGZdq{}}
\PYG{+w}{  }\PYG{c+cp}{\PYGZsh{}include}\PYG{+w}{ }\PYG{c+cpf}{\PYGZdq{}patch\PYGZus{}helpers.h\PYGZdq{}}
\PYG{+w}{  }
\PYG{+w}{  }\PYG{k+kt}{void}\PYG{+w}{ }\PYG{n+nf}{apply\PYGZus{}patch}\PYG{p}{(}\PYG{k+kt}{char}\PYG{+w}{ }\PYG{o}{*}\PYG{n}{rdi}\PYG{p}{,}\PYG{+w}{ }\PYG{k+kt}{char}\PYG{+w}{ }\PYG{o}{*}\PYG{n}{rsi}\PYG{p}{,}\PYG{+w}{ }\PYG{k+kt}{intptr\PYGZus{}t}\PYG{+w}{ }\PYG{o}{*}\PYG{n}{ret}\PYG{p}{,}\PYG{+w}{ }\PYG{k+kt}{intptr\PYGZus{}t}\PYG{+w}{ }\PYG{n}{size}\PYG{p}{)}
\PYG{+w}{  }\PYG{p}{\PYGZob{}}
\PYG{+w}{      }\PYG{k}{if}\PYG{+w}{ }\PYG{p}{(}\PYG{n}{size}\PYG{+w}{ }\PYG{o}{\PYGZlt{}=}\PYG{+w}{ }\PYG{l+m+mi}{1}\PYG{p}{)}\PYG{+w}{ }\PYG{p}{\PYGZob{}}
\PYG{+w}{          }\PYG{n}{basics\PYGZus{}set\PYGZus{}return}\PYG{p}{(}\PYG{n}{ret}\PYG{p}{,}\PYG{+w}{ }\PYG{p}{(}\PYG{k+kt}{intptr\PYGZus{}t}\PYG{p}{)}\PYG{n}{rdi}\PYG{p}{);}
\PYG{+w}{          }\PYG{k}{return}\PYG{p}{;}
\PYG{+w}{      }\PYG{p}{\PYGZcb{}}
\PYG{+w}{      }\PYG{n}{strncpy}\PYG{p}{(}\PYG{n}{rdi}\PYG{p}{,}\PYG{+w}{ }\PYG{n}{rsi}\PYG{p}{,}\PYG{+w}{ }\PYG{n}{size}\PYG{+w}{ }\PYG{o}{\PYGZhy{}}\PYG{+w}{ }\PYG{l+m+mi}{1}\PYG{p}{);}
\PYG{+w}{      }\PYG{n}{rdi}\PYG{p}{[}\PYG{n}{size}\PYG{+w}{ }\PYG{o}{\PYGZhy{}}\PYG{+w}{ }\PYG{l+m+mi}{1}\PYG{p}{]}\PYG{+w}{ }\PYG{o}{=}\PYG{+w}{ }\PYG{l+s+sc}{\PYGZsq{}\PYGZbs{}0\PYGZsq{}}\PYG{p}{;}
\PYG{+w}{      }\PYG{n}{basics\PYGZus{}set\PYGZus{}return}\PYG{p}{(}\PYG{n}{ret}\PYG{p}{,}\PYG{+w}{ }\PYG{p}{(}\PYG{k+kt}{intptr\PYGZus{}t}\PYG{p}{)}\PYG{n}{rdi}\PYG{p}{);}
\PYG{+w}{  }\PYG{p}{\PYGZcb{}}
\end{MintedVerbatim}
\endgroup
       \vspace{-3mm}
       \caption{Strcpy function patch template with statically determined buffer sizes.}
   \label{l:strcpy_patch}
   \vspace{-5mm}
\end{listing}

\begin{listing}[t]
    \begingroup\fvset{highlightcolor=gray!20,frame=lines, numbers=left, numbersep=-5pt, xleftmargin=0pt, fontsize=\footnotesize, framesep=2mm, breaklines}\begin{MintedVerbatim}[commandchars=\\\{\}]
\PYG{+w}{    }\PYG{c+cp}{\PYGZsh{}include}\PYG{+w}{ }\PYG{c+cpf}{\PYGZdq{}stdlib.c\PYGZdq{}}

\PYG{+w}{    }\PYG{k+kt}{void}\PYG{+w}{ }\PYG{n+nf}{apply\PYGZus{}patch}\PYG{p}{(}\PYG{k+kt}{bool}\PYG{+w}{ }\PYG{n}{canary}\PYG{p}{,}\PYG{+w}{ }\PYG{k+kt}{void}\PYG{+w}{ }\PYG{o}{*}\PYG{n}{rbp}\PYG{p}{,}\PYG{+w}{ }\PYG{k+kt}{char}\PYG{+w}{ }\PYG{o}{*}\PYG{n}{rdi}\PYG{p}{,}\PYG{+w}{ }\PYG{k+kt}{char}\PYG{+w}{ }\PYG{o}{*}\PYG{n}{rsi}\PYG{p}{)}
\PYG{+w}{    }\PYG{p}{\PYGZob{}}
\PYG{+w}{        }\PYG{k+kt}{long}\PYG{+w}{ }\PYG{n}{offset}\PYG{+w}{ }\PYG{o}{=}\PYG{+w}{ }\PYG{l+m+mi}{0}\PYG{p}{;}
\PYG{+w}{        }\PYG{k}{if}\PYG{+w}{ }\PYG{p}{(}\PYG{n}{canary}\PYG{p}{)}
\PYG{+w}{        }\PYG{p}{\PYGZob{}}
\PYG{+w}{            }\PYG{n}{offset}\PYG{+w}{ }\PYG{o}{=}\PYG{+w}{ }\PYG{l+m+mi}{8}\PYG{p}{;}
\PYG{+w}{        }\PYG{p}{\PYGZcb{}}
\PYG{+w}{        }\PYG{k+kt}{long}\PYG{+w}{ }\PYG{n}{size}\PYG{+w}{ }\PYG{o}{=}\PYG{+w}{ }\PYG{p}{(}\PYG{k+kt}{long}\PYG{p}{)}\PYG{n}{rbp}\PYG{+w}{ }\PYG{o}{\PYGZhy{}}\PYG{+w}{ }\PYG{p}{(}\PYG{k+kt}{long}\PYG{p}{)}\PYG{n}{rdi}\PYG{+w}{ }\PYG{o}{\PYGZhy{}}\PYG{+w}{ }\PYG{n}{offset}\PYG{p}{;}
\PYG{+w}{        }\PYG{n}{strncpy}\PYG{p}{(}\PYG{n}{rdi}\PYG{p}{,}\PYG{+w}{ }\PYG{n}{rsi}\PYG{p}{,}\PYG{+w}{ }\PYG{n}{size}\PYG{+w}{ }\PYG{o}{\PYGZhy{}}\PYG{+w}{ }\PYG{l+m+mi}{1}\PYG{p}{);}
\PYG{+w}{    }\PYG{p}{\PYGZcb{}}
\end{MintedVerbatim}
\endgroup
    \caption{\footnotesize Strcpy function patch template with buffer sizes determined at runtime.}
    \label{l:strcpy_patch_dynamic}
\end{listing}

\mypara{Patch Installation}
\tool installs patches by using the trampoline technique~\citep{e9patch} that allows binary rewriting. For each patchable
call-site, the patcher creates a rewriting rule that replaces the unsafe library
call with the selected \texttt{apply\_patch} payload. The rule passes the
arguments recovered at the original call site, including the relevant argument
registers, the return-value location, and the destination bound or runtime-bound
metadata. At runtime, execution reaches the original call site, is redirected to
the compiled payload, performs the bounded operation, and then resumes at the
instruction immediately following the replaced call.

Before reporting a patched binary, \tool checks that all requested rewriting
rules were installed. The patcher records the original and patched binary paths,
the patched call-sites, continuation addresses, selected templates, size mode,
bound metadata, and E9Patch installation result in a patch manifest. This
manifest is consumed by the validator and provides an auditable record of which
binary locations were rewritten. Because E9Patch can implement redirection
through trampoline or loader mappings, the original call may still be visible in
ordinary static disassembly; \tool therefore treats the installation result and
the patch manifest as evidence that the patch was successfully installed.

\mypara{Bounded Validation Evidence}
After rewriting, \tool executes the original and patched binaries on the same
validation cases. The case label is assigned before validation, based on where
the input came from. If the user provides a validation JSON file, each case uses
the file's \texttt{kind} field, which must be \texttt{malicious},
\texttt{benign}, or \texttt{boundary}. Otherwise, inputs produced by the
concolic execution phase for a violating path are treated as malicious. If no
such input is available, \tool creates a cyclic long-string fallback and treats
it as malicious. The validator then adds two preservation inputs: a short benign
string and a boundary string whose length is derived from the smallest recovered
destination bound.

Malicious cases test remediation: the original binary must expose the reported
vulnerable behavior, and the patched binary must avoid the corresponding crash,
timeout, or vulnerable-site failure. Benign and boundary cases test bounded
preservation: the patched binary must preserve the original exit code and
standard output under the selected comparison policy, and it must not introduce
a crash or timeout. Thus, the labels are not inferred from the patched result;
they define what the validator is supposed to check for that input.

The validator also checks whether the test input actually reaches the patched
call. It runs the patched binary under runtime observation, translates the
recorded patch address to the address used at runtime, records whether execution
reaches the patched call site, and captures the main stack-related registers
around the patched execution. This distinguishes a test that never reaches the
patched code from a test that runs through the patch and continues normally.

The validation report separates remediation from preservation. A remediation
pass means that the patched binary no longer shows the vulnerable behaviour
observed in the original execution. A preservation pass means that the tested
benign and boundary inputs produce the same observable behaviour, specifically
the same exit code and standard output, as the original binary. Otherwise, the
patch is classified as not behaviour preserving. This includes cases where the
patched execution changes observable behaviour, where the validation input does
not execute the patched code, or where the original binary already behaves
inconsistently across repeated executions of the same input, preventing a
reliable before-and-after comparison. These checks are limited to the executed
inputs; they provide bounded validation evidence for the installed patch, not a
proof of full functional equivalence.
\color{black}

\section{Implementation}
\label{sec:implementation}

\tool (\textit{Binary Analysis and Stack Integrity Checker System}) is
implemented in Python~3.11 for x86-64 Linux ELF binaries. It uses
angr~\citep{angr} for binary loading, disassembly, control-flow recovery, and
bounded symbolic execution; Spot~\citep{Spot} for LTL-to-Büchi translation;
E9Patch~\citep{e9patch} for binary rewriting; and GNU Debugger (GDB)~\citep{GNU1986} for runtime patch
validation. The implementation follows the four modules in
\Cref{fig:arch}: the Binary Data Extractor, the Security Property Converter,
the Model Checker, and the Vulnerability Patcher and Validator.
\tool is an open source tool, and it is available at~\citep{basics_github_a}.

\mypara{Binary Data Extractor}
The extractor loads the target binary with angr using
\texttt{auto\_load\_libs=False}. BASICS does not analyse external libraries as
part of the target program; instead, supported library calls are modelled by the
summary mechanisms described below. The extractor records binary metadata,
including the file path, file size, loader ranges, entry point,
sections, segments, and recovered functions. It then constructs the recovered
binary control-flow graph (bCFG) used by the rest of the pipeline.

BASICS supports two analysis modes. In whole-program mode, analysis starts from
the selected program entry point and explores the bCFG from that entry. This
mode preserves more calling context, but it can become expensive for large
binaries. In function-analysis mode, BASICS recovers the bCFG once, identifies
user functions with stack frames, and schedules each such function as an
independent analysis entry. Each function entry is checked using the same
MemStaCe construction, summary generation, and LTL model-checking code as
whole-program analysis. This mode allows BASICS to scale to large binary
collections by analysing functions separately, but its reports are treated as
candidate findings because caller context may be incomplete. Candidate reports
from function-analysis mode can then be rechecked from a containing caller or
program entry when confirmation is required.

\mypara{Security Property Converter}
Security properties are written as LTL formulas over the MemStaCe observations
defined in \Cref{sec:secProper}. The converter maps these formulas to BASICS
predicates, uses Spot to generate Büchi automata, and stores the automata in the
format used by the model checker. Generated automata are cached across runs.

\mypara{Model Checker}
The model checker constructs the \mss by traversing the recovered bCFG and
applying the stack-memory semantics from \Cref{sec:model_checker}. Each state
records the current program location, stack frames, byte states, buffer
metadata, canary metadata, and the current abstract memory action. Previously
processed combinations of program location and abstract stack state are reused
to avoid repeatedly expanding the same region.

Calls and loops are summarised before their effects are added to the \mss. For
supported C library routines, BASICS first recovers the call arguments, stack
destination, destination size, and affected byte range directly at the call
site. When local recovery is not enough, BASICS uses bounded symbolic execution
to summarise the call or loop: it searches from the current analysis entry to
the target, compares stack bytes before and after the operation, and converts
the differences into \mss updates. If the target is not reached within the
configured bounds, BASICS records the unresolved summary and continues from the
current abstract stack state.

After building the reachable \mss, the model checker evaluates the Büchi
automata generated from the LTL properties. If a property is violated, BASICS
reports the violated property, the counterexample trace, and any concrete inputs
generated during summary construction. Otherwise, it reports the properties that
hold over the explored \mss.

\mypara{Vulnerability Patcher and Validator}
The patcher uses counterexample traces to identify supported vulnerable library
calls and select matching patch templates. The current patch set covers
\texttt{strcpy}, \texttt{gets}, \texttt{scanf}, \texttt{fscanf},
\texttt{sscanf}, \texttt{strcat}, \texttt{sprintf}, \texttt{memcpy}, and
\texttt{memmove}. Templates either use a recovered destination size or compute a
runtime bound from the stack frame.

Larger routines shared by multiple templates are factored into
\texttt{patch\_helpers.h}. These helpers implement bounded token reads, bounded
string extraction, the supported \texttt{scanf}/\texttt{fscanf}/\texttt{sscanf}
conversions, bounded \texttt{gets}-style line reading, and uniform return-value
handling. Keeping this logic in a helper header keeps individual templates
small while preserving the same guarantee: helpers write at most the recovered
capacity, terminate string destinations when applicable, and return a value
compatible with the replaced library routine.

BASICS installs selected patches with E9Patch~\citep{e9patch}. For each patched call site, it
records the selected template, recovered or runtime bound, continuation address,
and E9Patch installation result in a patch manifest. A binary is reported as
patched only when all requested call sites are successfully rewritten.

After patch installation, the validator runs the original and patched binaries
on the same malicious, benign, and boundary inputs. Malicious inputs check
whether the patch removes the observed crash or vulnerable behaviour. Benign and
boundary inputs check that the patched binary preserves the original exit code
and standard output and does not introduce a crash or timeout.

During this runtime validation stage, BASICS also uses GDB to observe the
patched execution. For each validation input, GDB starts the patched binary,
resolves the runtime address of the patched call site, sets a breakpoint at that
site, and records whether execution reaches the patch. When the breakpoint is
hit, BASICS records the main stack-related registers around the patched call.
The final report separates vulnerability remediation from bounded
functional preservation. A patch is marked as remediated when the malicious
input no longer triggers the observed crash or property violation. It is marked
as behaviour preserving only when the patched binary keeps the original exit code
and standard output on benign and boundary inputs without introducing a crash
or timeout. Otherwise, the patch is classified as not behaviour preserving,
including cases where the validation input does not reach the patched call site,
the original execution does not provide a stable baseline, or the patched
execution produces different observable behaviour.

\section{Experimental Evaluation}
\label{sec:evaluation}

We evaluate \tool along five dimensions: the fidelity of the generated \mss, the effectiveness of the LTL security properties as vulnerability
specifications, the ability of the model checker to detect BOs, the effectiveness of generated patches, and the scalability of the
implementation on real binaries. This leads to the following research questions:
\textbf{Q1.} Does \tool detect stack-based buffer overflows with high precision?
\textbf{Q2.} How does \tool compare with existing analysis tools on the same
datasets?
\textbf{Q3.} Are generated patches effective on patchable vulnerabilities?
\textbf{Q4.} Can \tool be applied to real open-source programs and Linux package
binaries?
\textbf{Q5.} What are the main causes of false positives (FPs) and false negatives (FNs)?

\subsection{Evaluation Setup and Grading Protocol}

\subsubsection{Evaluation Setup}
The evaluation is organised into four parts. First, we measure detection
accuracy and execution performance on controlled benchmark datasets with known vulnerable and non-vulnerable cases. Second, we evaluate binary patch generation and bounded
patch validation on patchable programs from one of the controlled datasets. Third, we run \tool on larger open-source C/C++ projects to assess scalability under realistic program structure. Finally, we perform a Linux package sweep to evaluate whether the function-by-function analysis mode can be scheduled across many independent real-world binaries.

For controlled benchmarks, we used the Juliet C/C++ test suite~\citep{nist-juliet}
and NIST SARD~\citep{nist-sard}.  The Juliet subset contains 3,524 isolated
CWE-121 binaries, with 1,762 vulnerable and 1,762 non-vulnerable cases.  The
SARD subset contains 151 C programs, with 51 vulnerable and 100
non-vulnerable cases.
\DIFF{For open-source projects and Linux packages, we used 5 (3 and 2, respectively) that contain 292 binaries. For Linux packages, we specifically chose the Core Linux package and the Additional packages associated with it, which together comprise 58 packages.}

All experiments were run on Ubuntu 24.04.4 LTS on a machine with 48 AMD EPYC
7643 hardware threads and 251 GiB of RAM.  The benchmark harness schedules
independent cases in parallel and records per-case status, elapsed time,
reported vulnerabilities, and confusion-matrix data.  Unless stated
otherwise, the reported time is the sum of per-case elapsed times; for the
large sweeps, we also report wall-clock time measured from the first case start
to the last case finish.

\subsubsection{Grading Protocol}

A binary is classified as vulnerable by \tool if the
model checker reports at least one violation of the stack-security properties
defined in \Cref{fig:basics_ltl_properties}.
Since for Juliet and SARD, the benchmark labels provide ground truth, each report can be classified as a true positive or false positive at the benchmark-case level. For open-source and Linux package sweeps, since they do not provide such labels, reports from these sweeps are therefore treated as candidate vulnerabilities until they are confirmed by follow-up manual review.

Additionally, using Juliet and SARD, we compare \tool with four reproducible tools that can be applied directly to automated vulnerability analysis, since other tools discussed later in \Cref{sec:rw} either do not provide publicly available implementations or require manually prepared targets and inputs. 
Hence, for \tool comparison, we chose the tools CWE\_Checker~\citep{cwe_checker}, Flawfinder~\citep{flawfider}, Manticore~\citep{manticore}, and BinAbsInspector~\citep{binabsinspector}. These baselines are not all dedicated binary-level stack-buffer-overflow classifiers: Flawfinder is source-code only,
Manticore is a general symbolic-execution engine, and CWE\_Checker and
BinAbsInspector combine memory-safety checks with broader
dangerous-function diagnostics. For tools that provide both direct
memory-safety reports and unsafe-function warnings, we include both
classes in the comparison. Consequently, CWE\_Checker and
BinAbsInspector are evaluated with CWE-676 dangerous-function checks
enabled in addition to stack memory-safety reports. Since CWE-676
represents unsafe API usage rather than direct evidence of stack
corruption, this configuration can increase FPs. 

In this comparison, positive reports on vulnerable benchmark-cases are true positives, while positive reports on non-vulnerable benchmarks are FPs. 
Specifically, \tool is graded positive when the model checker reports a violation of
the stack-security LTL properties. CWE\_Checker is graded positive when
its JSON output contains CWE-119, CWE-787, or CWE-676. Flawfinder is
graded positive when its CSV output contains a buffer-overflow CWE-family
report, excluding the generic static \texttt{char} array rule. Manticore
is graded positive only when generated test cases trigger a crash,
segmentation fault, or invalid-memory access. BinAbsInspector is graded
positive when it reports CWE-676, stack-relevant CWE-119/CWE-787, or
stack out-of-bounds violations; heap out-of-bounds reports are filtered
because the benchmark targets stack vulnerabilities.

For tools with failed benchmark cases, metrics are computed only over completed
cases; \emph{Errors} are reported in cases of timed out, failed to build or could not be run, and therefore excluded from the confusion matrix.

\begin{table*}[t]
\caption{\small Detection results on the SARD dataset.}
\label{tab:sard-detection}
\vspace{-2mm}
\scriptsize
\centering
\resizebox{0.85\textwidth}{!}{%
\begin{tabular}{@{}lcccccrrrrr@{}}
\toprule
\textbf{Tool} & \textbf{Errors} & \textbf{TP} & \textbf{FP} & \textbf{TN} & \textbf{FN} &
\textbf{Precision} & \textbf{Recall} & \textbf{F1} & \textbf{Accuracy} & \textbf{Sum time} \\ \midrule
\tool & 0 & 38 & 6 & 94 & 13 & 86.4\% & 74.5\% & 80.0\% & 87.4\% & 13m 0s \\
BinAbsInspector & 0 & 0 & 0 & 100 & 51 & N/A & 0.0\% & 0.0\% & 66.2\% & 1h 2m 7s \\
CWE\_Checker & 0 & 42 & 70 & 30 & 9 & 37.5\% & 82.4\% & 51.5\% & 47.7\% & 59m 9s \\
Flawfinder & 0 & 48 & 93 & 7 & 3 & 34.0\% & 94.1\% & 50.0\% & 36.4\% & 11.8s \\
Manticore & 27 & 16 & 9 & 75 & 24 & 64.0\% & 40.0\% & 49.2\% & 73.4\% & 1h 0m 9s \\ \bottomrule
\end{tabular}}
\end{table*}

\begin{table*}[t]
\caption{\small Detection results on the Juliet CWE-121 dataset.}
\label{tab:juliet-detection}
\vspace{-2mm}
\scriptsize
\centering
\resizebox{0.85\textwidth}{!}{%
\begin{tabular}{@{}lcccccrrrrr@{}}
\toprule
\textbf{Tool} & \textbf{Errors} & \textbf{TP} & \textbf{FP} & \textbf{TN} & \textbf{FN} &
\textbf{Precision} & \textbf{Recall} & \textbf{F1} & \textbf{Accuracy} & \textbf{Sum time} \\ \midrule
\tool & 0 & 900 & 145 & 1,617 & 862 & 86.1\% & 51.1\% & 64.1\% & 71.4\% & 14h 40m 41s \\
BinAbsInspector & 0 & 8 & 36 & 1,726 & 1,754 & 18.2\% & 0.5\% & 0.9\% & 49.2\% & 29h 39m 37s \\
CWE\_Checker & 0 & 1,472 & 1,454 & 308 & 290 & 50.3\% & 83.5\% & 62.8\% & 50.5\% & 32h 39m 21s \\
Flawfinder & 0 & 1,108 & 1,090 & 672 & 654 & 50.4\% & 62.9\% & 56.0\% & 50.5\% & 5m 38s \\
Manticore & 101 & 671 & 0 & 1,712 & 1,040 & 100.0\% & 39.2\% & 56.3\% & 69.6\% & 29h 23m 5s \\ \bottomrule
\end{tabular}}
\end{table*}

\subsection{Detection on SARD and Juliet Subsets}
\label{sec:detection}

\subsubsection{SARD}

Table~\ref{tab:sard-detection} shows the SARD results. \tool completed all
151 cases and achieved the best accuracy and F1-score among the five tools.
Its precision was 86.4\%, with 6 FPs and 13 FNs. Flawfinder reached a higher recall, but at the cost of a large number
of FPs. CWE\_Checker detected the largest number of vulnerable cases
among the binary tools, but most of its positive signals came from CWE-676
dangerous-function warnings rather than direct stack-corruption reports. This
improves recall, but also produces 70 FPs because unsafe API use is
only a candidate vulnerability signal. BinAbsInspector produced no positives on
SARD, even with CWE-676 enabled, giving perfect specificity but zero recall.
Manticore produced useful results on the cases it completed, but timed out on
27 out of the 151 cases.

\tool also completed the benchmark substantially faster than the binary-analysis baselines. Its aggregated analysis time was 13m 0s, compared to approximately one hour for CWE Checker, BinAbsInspector, and Manticore. Flawfinder completed faster still, but operates directly on source code rather than binary state exploration.

These results answer Q1. positively on small and medium-sized benchmark
programs: when \tool constructs a sufficiently precise state space, property
violations usually correspond to real stack corruption. The remaining FNs are mainly caused by imprecise modelling of library functions and by
state-space truncation in loops. The FPs occur when conservative
stack-state approximations mark frame metadata as modified even though the
source program uses bounded operations.

\subsubsection{Juliet}

Table~\ref{tab:juliet-detection} summarizes the Juliet CWE-121 results. The
Juliet dataset is substantially harder for \tool because many cases contain
deeper call chains, loop-dependent writes, and library functions whose effects
must be approximated. \tool still preserves high precision, 86.1\%, but its
recall drops to 51.1\%. This indicates that reported violations are usually
meaningful, while missed cases are primarily due to incomplete state-space
construction or imprecise function modelling rather than to weak security
properties.

CWE\_Checker detected the largest number of vulnerable cases among the
binary-analysis tools, but this came with a high FP rate because
many safe Juliet binaries still call APIs classified as dangerous.
BinAbsInspector had the opposite behavior: it produced few reports, so its
FP rate was low, but it missed nearly all vulnerable cases.
Flawfinder finds more vulnerable cases than \tool but also reports many
non-vulnerable programs as vulnerable, i.e., FPs. Manticore achieved high precision on the completed subset, but timed out on 101 cases.

Runtime differences were more pronounced on Juliet. \tool completed the full
benchmark in 14h 40m, while CWE Checker, BinAbsInspector, and Manticore each
required approximately 29--33 hours of aggregated analysis time. Flawfinder was
substantially faster because it performs lightweight source-level pattern
matching rather than binary-level state exploration and verification.

\DIFF{We further categorised the 862 false negatives produced by \tool on Juliet according to their dominant cause. Approximately 44\% (379 cases) resulted from incomplete modelling of copy, formatting, inlined-write, or wide-character operations. Approximately 37\% (319 cases) arose from missing execution context or bounded exploration, including runtime-dependent inputs, and loops. The remaining 19\% (164 cases) were source-level off-by-one cases where the out-of-bounds write fell into compiler-added padding or alignment space. As a result, the write did not overwrite an adjacent stack object in the compiled binary.}

\subsubsection{Results Discussion on Both Subsets}

\begin{table}[t]
\caption{\small Case-level detection overlap between \tool and CWE\_Checker}
\label{tab:detection-overlap}
\vspace{-2mm}
\centering
\small
\begin{tabular}{@{}lrrrr@{}}
\toprule
\textbf{Dataset} & \textbf{Both} & \textbf{\tool only} &
\textbf{CWE\_Checker only} & \textbf{Neither} \\
\midrule
SARD   & 30  & 8   & 12  & 1   \\
Juliet & 794 & 106 & 678 & 184 \\
\bottomrule
\end{tabular}
\end{table}

\DIFF{Since \tool and CWE\_Checker were the two best tools on both subsets, a further analysis was done to understand where they overlap or do not overlap in vulnerable case detection.
Table~\ref{tab:detection-overlap} compares both tools at the case level, considering only ground-truth vulnerable cases. On SARD, both tools detected the same 30 cases, while 8 were detected only by \tool and 12 only by CWE\_Checker. On Juliet, both tools detected 794 of the same cases, while 106 were unique to \tool and 678 to CWE\_Checker. Although most of \tool's true positives overlap with those of CWE\_Checker, \tool still identifies vulnerable cases that CWE\_Checker misses.} 

Together, the SARD and Juliet results answer Q2. \tool provides the strongest overall balance between precision, recall, accuracy, and analysis cost among the evaluated tools. The comparison also highlights important differences between analysis approaches: source-level pattern matching, binary dangerous-function diagnostics, symbolic test generation, and LTL-based binary model checking expose different forms of evidence and therefore produce different precision-recall tradeoffs. \tool's main limitation is recall on complex Juliet programs, which is consistent with the cost of path exploration, loop handling, and function emulation in binary-level symbolic execution.

\subsection{Patching on SARD}

We also evaluated patch generation and bounded patch validation on SARD. Each
patch records a manifest and is validated over generated or provided malicious,
benign, and boundary inputs. The Validator reports vulnerability-mitigation
evidence separately from bounded functional-preservation evidence, and does not
claim full functional preservation for the patching process.

\begin{table}[b]
\caption{\small SARD benchmark composition and structured patch validation.}
\label{tab:sard-patching}
\vspace{-2mm}
\centering
\resizebox{\columnwidth}{!}{%
\begin{tabular}{@{}lccccc@{}}
\toprule
\textbf{Class} & \textbf{Cases} & \textbf{Vuln.} & \textbf{Patched} &
\textbf{Passed} & \textbf{Failed Preservation} \\ \midrule
\texttt{gets} & 2 & 2 & 2 & 2 & 0 \\
\texttt{fscanf} & 7 & 4 & 4 & 4 & 0 \\
\texttt{scanf} & 16 & 6 & 6 & 6 & 0 \\
\texttt{sscanf} & 3 & 1 & 2 & 2 & 0 \\
\texttt{strcpy} & 51 & 19 & 15 & 14 & 1 \\
\texttt{strcpy} underwrite & 5 & 5 & 5 & 5 & 0 \\
\texttt{sprintf} & 18 & 13 & 12 & 8 & 4 \\
\texttt{vsprintf} & 4 & 1 & 0 & 0 & 0 \\
Other not-vulnerable & 45 & 0 & 0 & 0 & 0 \\ \midrule
\textbf{Total} & \textbf{151} & \textbf{51} & \textbf{46} &
\textbf{41} & \textbf{5} \\ \bottomrule
\end{tabular}}
\end{table}

\begin{table*}[t]
\caption{\small Average per-case pipeline-stage runtime from a separate
instrumented profiling experiment. SARD uses all 151 cases, while Juliet uses
a stratified 200-case subset containing 100 matched bad/good pairs. Values are
arithmetic mean $\pm$ standard deviation in seconds. End-to-end time includes
the measured stages and unattributed runtime overhead.}
\label{tab:stage-runtime}
\vspace{-2mm}
\centering
\resizebox{0.9\textwidth}{!}{%
\begin{tabular}{@{}lrrrrrrr@{}}
\toprule
\textbf{Dataset} &
\textbf{Disassembly/bCFG} &
\textbf{MemStaCe} &
\textbf{LTL checking} &
\textbf{Patch generation} &
\textbf{Patch Validation} &
\textbf{End-to-end} \\ \midrule
SARD &
 $0.078 \pm 0.005$s &
  $0.052 \pm 0.277$s &
  $0.048 \pm 0.024$s &
  $0.006 \pm 0.009$s &
  $1.028 \pm 2.027$s &
  $1.338 \pm 2.050$s \\
Juliet &
$0.244 \pm 0.012$s &
$13.741 \pm 42.646$s &
$0.211 \pm 0.146$s &
N/A &
N/A &
$14.317 \pm 42.657$s \\ \bottomrule
\end{tabular}}
\end{table*}

Table~\ref{tab:sard-patching} shows both the functions and patterns represented
in the SARD subset and the patch-validation outcome. The dataset contains 151
cases, of which 51 contain BO vulnerabilities. The Patcher produced 46
patched binaries. 41 patched binaries passed both vulnerability
mitigation and bounded functional-preservation checks. The remaining 5 patched binaries removed the observed vulnerable behavior but failed bounded
preservation: the patched executions differed from the original executions on
the validator's benign or boundary inputs.

The input-function classes were the most direct cases. Both \texttt{gets}
cases, all four patched \texttt{fscanf} cases, all six patched \texttt{scanf}
cases, and both patched \texttt{sscanf} cases passed validation. The
\texttt{strcpy} class produced 15 patched binaries; 14 passed validation and
one failed preservation. All 5 \texttt{strcpy} underwrite cases passed
validation. The \texttt{sprintf} class produced 12 patched binaries; 8
passed validation and 4 failed preservation. These failures occurred in
format-sensitive executions where the patch removed the unsafe write but
changed observable behavior, such as formatted output or termination behavior.
No patch was produced for the vulnerable \texttt{vsprintf} case.

These results answer Q3. positively within the bounds of the validator. The generated patches removed the observed vulnerable behavior in all patched binaries, and 41 out of the 46 patched binaries preserved the original exit code and standard output on the validator's benign and boundary inputs. The remaining 5 cases changed observable behavior after patching, primarily in
format-sensitive functions such as \texttt{sprintf}.

\subsection{Pipeline Runtime Profiling on Ground-truth Subsets}
\label{ssec:runtime-profiling}

\DIFF{We conducted a separate profiling experiment to complement the aggregated
runtime results reported in Section~\ref{sec:detection},~Tables~\ref{tab:sard-detection}
and~\ref{tab:juliet-detection}. This experiment instrumented \tool to measure
the elapsed time spent in binary disassembly and bCFG recovery, \mss
construction, LTL model checking, patch generation, and patch validation.
It uses all 151 SARD cases and a stratified
200-case Juliet subset containing 100 matched bad/good pairs.
Table~\ref{tab:stage-runtime} reports the average and standard deviation
of the per-case elapsed time for each stage. End-to-end time includes the
measured stages and unattributed pipeline overhead.}

\DIFF{\mss construction dominates the Juliet profiling results, accounting
for 13.741\,s of the 14.317\,s average end-to-end runtime. For SARD, patch
validation is the largest component, averaging 1.028\,s of the 1.338\,s
end-to-end runtime, while disassembly, LTL checking, and patch generation each
remain below 0.1\,s on average. Patch generation and validation are not
reported for Juliet because the Juliet profiling experiment evaluates detection
only.}

\begin{table*}[t]
\caption{\small Open-source project sweep and function-level candidate
filtering. \emph{Filtered (\%)} reports the percentage of initial candidates
removed, and \emph{Cand. bins} counts binaries containing at least one final
candidate.}
\label{tab:oss-projects}
\vspace{-2mm}
\centering
\resizebox{\textwidth}{!}{%
\begin{tabular}{@{}lrrrrrrrrrrrr@{}}
\toprule
\textbf{Project} &
\textbf{LoC} &
\textbf{Binaries} &
\textbf{Errors} &
\textbf{Funcs} &
\textbf{Initial cand.} &
\textbf{Filtered} &
\textbf{Final cand.} &
\textbf{Filtered (\%)} &
\textbf{Cand. bins} &
\textbf{Sum time} &
\textbf{Wall time} &
\textbf{Total bin. size} \\ \midrule
cJSON &
22,383 & 23 & 0 & 2,537 &
0 & 0 & 0 & -- & 0 &
7h 16m 25s & 57m 21s & 2.2 MiB \\
libplctag &
95,649 & 25 & 4 & 12,440 &
0 & 0 & 0 & -- & 0 &
5h 12m 26s & 57m 0s & 172.3 MiB \\
SOEM &
25,305 & 7 & 2 & 736 &
7 & 1 & 6 & 14.3\% & 1 &
1h 14m 28s & 30m 4s & 1.9 MiB \\ \midrule
\textbf{Total} &
\textbf{143,337} &
\textbf{55} &
\textbf{6} &
\textbf{15,713} &
\textbf{7} &
\textbf{1} &
\textbf{6} &
\textbf{14.3\%} &
\textbf{1} &
\textbf{13h 43m 19s} &
\textbf{2h 24m 25s} &
\textbf{176.4 MiB} \\ \bottomrule
\end{tabular}}
\end{table*}

\begin{table*}[t]
\caption{\small Linux package sweep and function-level candidate filtering.
\emph{Filtered (\%)} reports the percentage of initial candidates removed, and
\emph{Cand. bins} counts binaries containing at least one final candidate.}
\label{tab:linux-sweep}
\vspace{-2mm}
\centering
\resizebox{\textwidth}{!}{%
\begin{tabular}{@{}lrrrrrrrrrrr@{}}
\toprule
\textbf{Sweep} &
\textbf{Packages} &
\textbf{Binaries} &
\textbf{Funcs} &
\textbf{Initial cand.} &
\textbf{Filtered} &
\textbf{Final cand.} &
\textbf{Filtered (\%)} &
\textbf{Cand. bins} &
\textbf{Sum time} &
\textbf{Wall time} &
\textbf{Total bin. size} \\ \midrule
Core Linux packages &
21 & 117 & 1,737 &
38 & 38 & 0 & 100.0\% & 0 &
14h 10m 54s & 1h 50m 3s & 32.5 MiB \\
Additional packages &
37 & 120 & 4,016 &
221 & 220 & 1 & 99.5\% & 1 &
11h 12m 18s & 1h 28m 35s & 14.0 MiB \\ \midrule
\textbf{Combined} &
\textbf{58} &
\textbf{237} &
\textbf{5,753} &
\textbf{259} &
\textbf{258} &
\textbf{1} &
\textbf{99.6\%} &
\textbf{1} &
\textbf{25h 23m 12s} &
\textbf{3h 18m 47s} &
\textbf{46.6 MiB} \\ \bottomrule
\end{tabular}}
\end{table*}

\subsection{Scalability on Open-Source Projects}

To evaluate the \tool's scalability on real programs, we built and scanned three larger
open-source C/C++ projects: cJSON\footnote{\url{https://github.com/DaveGamble/cJSON}},
libplctag\footnote{\url{https://github.com/libplctag/libplctag}}, and
SOEM\footnote{\url{https://github.com/OpenEtherCATsociety/SOEM}}.
Because whole-binary analysis can exhaust memory on larger binaries, we used
function-by-function analysis. \tool enumerates functions with local stack
frames, analyses each function as an entry point, and records whether the
analysis is clean or produces a candidate report.

\DIFF{The evaluated projects cover 143,337 lines of C/C++ code, 55 binaries,
and 15,713 analysed function entries. Table~\ref{tab:oss-projects} summarises
the analysis results, candidate filtering, and runtime measurements. The initial
function-by-function sweep produced 7 candidate functions, all in SOEM. The
isolated RBP-integrity filter removed one candidate (14.3\%), leaving 6
functions in one binary for contextual review.}

\DIFF{The isolated RBP-integrity filter removes candidate functions whose only
evidence is a saved-frame-pointer integrity violation, without an accompanying
stack-boundary, return-address, canary, underwrite, or unsafe-write violation.
In function-by-function mode, such reports are frequently caused by missing
caller context or conservative stack-state modeling and are therefore excluded
from contextual review.}

A focused re-analysis of the six final SOEM candidate functions did not
confirm a real vulnerability. Two SOEM binaries and four libplctag binaries did
not complete analysis. Five cases reached the 1,800\,s analysis timeout, while
one SOEM binary terminated because of an out-of-memory error during symbolic
execution. The remaining binaries completed successfully with an aggregate
analysis time of 13h 43m and a wall-clock time of 2h 24m under parallel
execution.

Across 55 binaries and 15,713 analysed functions, the OSS sweep produced only a
small number of candidate reports while remaining tractable in aggregate
wall-clock time. These results support Q4. by showing that function-by-function
analysis allows \tool to scale to medium-to-large scale real-world binaries without requiring whole-program verification.

\subsection{Function-by-function Analysis on Linux Package Sweep}

We also evaluated \tool's function-by-function analysis ability on binaries extracted from Linux packages.  This sweep was designed to test breadth: many small and medium binaries from common command-line packages, followed by an additional set of less prominent packages that process structured inputs such as archives, images, documents, and network data.

\DIFF{As shown in Table~\ref{tab:linux-sweep}, the Linux sweep covered 58
packages, 237 binaries, and 5,753 analysed functions across a diverse set of
real-world Linux utilities. The initial function-by-function sweep produced 259
candidate functions. Applying the same isolated RBP-integrity filter removed
258 candidates (99.6\%), leaving one function in one binary for contextual
review.}

The low candidate rate suggests that the filtering and caller-confirmation pipeline substantially reduces the review burden introduced by
function-by-function analysis. Follow-up validation did not confirm the Linux
candidate as a real vulnerability. These results support Q4. by showing that
\tool can be scheduled across hundreds of real Linux binaries while remaining
tractable in terms of total execution time and the volume of candidate-review.

\subsection{Candidate Review}

When applied to large real-world binaries, \tool operates in two stages. First,
it performs a scalable function-by-function sweep, analysing each recovered
function independently and flagging functions that violate one or more security
properties as vulnerability candidates. This initial stage prioritizes coverage
and scalability, allowing large binaries to be analysed without requiring
whole-program exploration. Second, each candidate that remains after the isolated RBP-integrity filter
undergoes contextual validation, where the analysis is repeated using additional
execution context, typically from the program entry point or from a direct
caller function. A candidate is considered confirmed only if the same violated property reappears
and the counterexample trace still reaches the originally reported function.

The open-source and Linux sweeps produced six and one final candidate functions
after filtering, respectively. However, follow-up contextual validation did not confirm any
of these candidates as genuine stack-based buffer overflow vulnerabilities.
Consequently, no previously unknown vulnerabilities were identified in the
evaluated software. Candidate reports were re-analysed from the program entry
point when the candidate function was itself the analysis entry, and from direct
caller functions otherwise.

The SOEM \texttt{slaveinfo} candidates were produced by isolated
function-by-function analysis. The flagged functions include formatting helpers
such as \texttt{dtype2string}, \texttt{otype2string},
\texttt{access2string}, and \texttt{SDO2string}, plus mapping routines such as
\texttt{si\_PDOassign} and \texttt{si\_map\_sdo}. Manual inspection showed
that these functions use bounded formatting operations, for example
\texttt{snprintf} with \texttt{sizeof}-bounded static buffers, rather than
unchecked writes to stack buffers. They were flagged because isolated analysis
starts these functions without their real caller state and with highly
unconstrained inputs. In addition, some EtherCAT helper calls and loops could
not be fully emulated, which caused conservative stack-state approximations and
erroneous underflow or frame-integrity violations.

The Linux sweep produced one candidate in \texttt{net-tools}/\texttt{netstat},
reported at \texttt{sub\_40ee38}. This was not a real function entry but an
internal basic block inside a larger function. \tool flagged it because the
function-recovery heuristic treated the block as an independent function and
analysed it without the prologue and caller context that establish the stack
frame. Re-running the analysis on the containing function removed the
violation. Thus, this candidate was a control-flow recovery artifact rather
than a memory-safety bug.

These results answer Q5. Function-by-function analysis enables \tool to scale
to large binaries by reducing the cost of whole-program exploration. As
expected, analysing functions in isolation may introduce imprecision when
caller context, function boundaries, external-call effects, or loop behaviour
cannot be fully reconstructed. The contextual validation stage eliminated all
final candidates in our evaluation, demonstrating its effectiveness as a
filtering step. Nevertheless, the results indicate that improvements in
caller-context reconstruction and function-boundary recovery could further
reduce the number of candidates requiring follow-up analysis in large
real-world binaries.

\section{Discussion and Limitations}
\label{sec:lim}

\mypara{Threat Model and Memory Scope}
\tool is designed to detect stack-based memory-safety violations whose effects
can be represented in the current \mss. The model focuses on stack objects,
stack-frame metadata, and writes that may corrupt adjacent stack state. This is
sufficient for detecting stack-based buffer overflows and corruption of
control-relevant stack data, but it does not cover vulnerabilities whose effects
are confined to other memory regions, such as heap overflows, use-after-free, or
double-free bugs. In addition, the current memory model mainly tracks writes and
does not explicitly model memory reads. As a result, \tool does not claim to
detect every memory-safety or control-flow attack, but rather stack-memory violations that manifest as violations of the modelled security properties.

Extending the \mss with heap objects, read tracking, and richer control-flow
metadata would allow \tool to express more complex security properties,
including properties for heap vulnerabilities and attacks that depend on
read/write interactions or control-flow hijacking. These extensions are outside
the scope of the current work.

\mypara{Bounded Analysis}
\tool is a bounded analysis. It limits loop exploration, recursion depth, and
symbolic execution in order to keep verification tractable. As a result,
vulnerabilities that require deeper exploration may be missed. This behaviour is
visible in the Juliet evaluation, where deeper call chains and loop-dependent writes contributed to false negatives.

The analysis also depends on approximating the behaviour of loops and external
library functions. In some cases, these approximations are too conservative and report writes that are not actually possible, while in other cases they miss
relevant memory updates. Since \tool builds on angr, improvements in angr's
binary analysis and symbolic execution directly improve the precision of the
generated memory-state space.

\mypara{Patch Validation}
\tool provides automated patch generation together with automated bounded
validation, but it does not prove full behavioural equivalence between the
original and patched binaries. The Validator checks whether malicious inputs no longer trigger the observed vulnerable behaviour and whether selected benign and
boundary inputs preserve the original exit code and standard output.

This provides bounded evidence that a patch removes the observed vulnerability
without breaking the tested executions, but it does not guarantee that all
program behaviours are preserved. Stronger validation would require a larger input
corpora or equivalence-checking techniques, which are outside the scope of this
work.

\mypara{Compatibility with Existing Defenses}
\tool operates on recovered binary control flow and stack-memory state, and is
therefore compatible with common mitigation mechanisms such as stack canaries,
ASLR, and DEP/NX during vulnerability detection. In particular, BASICS
explicitly models stack-canary integrity as one of its security properties.
However, the current patching approach relies on trampoline-based binary
rewriting through E9Patch. As a result, strict control-flow integrity (CFI) enforcement may restrict or invalidate some runtime patch redirections,
depending on the target environment and runtime policy.

\mypara{Scalability and Caller Context}
The main scalability limitation of \tool is full-binary analysis. For larger
programs, analysis starting from the program entry point can exhaust memory or
hit the analysis timeout. To address this, \tool supports function-by-function
analysis, where functions are analysed independently.

This mode improves scalability, but it also removes part of the caller context.
As a result, function arguments, pointer relationships, and stack state may be
modelled imprecisely, which can produce false positives. The real-binary sweeps
showed that these issues can arise from incomplete caller information,
incorrectly recovered function boundaries, or conservative stack modelling.
Future work should improve caller-context reconstruction and function recovery
for large binaries.

\section{Extensibility of \tool}
\label{sec:extension}

\tool was designed to be extensible by users, with support for user-provided
security properties, patch templates, and CWE vulnerability maps. The \mss is
generic enough to support other security-property checks over stack memory.

To verify custom security properties, users can write LTL formulas using the
standard LTL operators and the query functions from
\Cref{fig:memstace_query_functions} to inspect the \mss model. Users may also
implement additional query functions when the existing ones are insufficient.
The tool then automatically compiles the LTL formula and uses it in subsequent
verification runs.

By default, user-defined security properties are not associated with any known
vulnerability class. To classify them, \tool uses a CWE vulnerability map that
associates security properties with CWE classes. When new properties are added,
users can map them to existing CWEs or add new map entries. This allows \tool
to report property violations using the corresponding CWE classification.

If a newly reported vulnerability should also be repaired, the patching module
requires a corresponding patch template for the identified sink. Users can
extend the set of supported sinks by providing new patch templates. These
templates are written as small C code snippets that include the custom
\texttt{stdlib.c} file provided by E9Patch. Each template must define a
\texttt{void} function named \texttt{apply\_patch}, 
which receives as arguments the registers used to pass
parameters according to the System V calling convention.

Finally, since \tool is open source~\citep{basics_github_a}, users can modify
the tool directly to add support for additional architectural or memory-model
features. For example, one could model the x86-64 red zone by extending the
stack-frame model with the 128-byte region below the stack pointer.

\section{Related Work}
\label{sec:rw}

Prior work relevant to BASICS spans three largely separate areas: binary-level
vulnerability discovery, formal verification of program safety properties, and
automated repair. Existing binary-analysis systems are effective at discovering
bugs or providing runtime bug oracles, while formal verification and automated
repair systems have more commonly been applied at the source-code level. BASICS combines these threads in a narrower setting: detecting stack-based buffer overflows in x86-64 binaries through explicit stack-memory properties and using the resulting counterexamples to guide binary patching.
This section positions BASICS against recent binary-analysis and sanitisation
systems, source-level model checkers and repair techniques, and binary-rewriting
systems.

\newcolumntype{Y}{>{\raggedright\arraybackslash}X}

\begin{table*}[!ht]
\centering
\scriptsize
\setlength{\tabcolsep}{4pt}
\renewcommand{\arraystretch}{1.12}
\caption{Comparison of BASICS with representative binary-level vulnerability-analysis systems.}
\label{tab:binary-tools-comparison}
\begin{tabularx}{\textwidth}{@{}p{2.0cm}p{2.65cm}p{3.15cm}Y Y@{}}
\toprule
\textbf{System} &
\textbf{Target} &
\textbf{Method} &
\textbf{Primary contribution} &
\textbf{Scope relative to BASICS} \\
\midrule

\textbf{BASICS} &
Stack BOs in x86-64 Linux binaries &
Bounded model checking of stack-memory properties with LTL; bounded concolic execution for selected calls and loops; binary rewriting for supported sinks &
Links stack-property violations to bounded validation and binary patching &
Focused on stack BOs; patching limited to supported sinks and recovered stack objects \\
\midrule

\textbf{CWE\_Checker}~\citep{cwe_checker} &
Several CWE classes in binaries &
Static binary analysis over a lifted IR &
Practical multi-CWE binary triage &
Detection and warning generation; no LTL stack model or automated repair \\
\midrule

\textbf{ZAFL}~\citep{zfal} &
Crash discovery in binary-only programs &
Binary instrumentation for coverage-guided fuzzing &
Improves fuzzing speed and coverage without source code &
Optimises bug discovery through crashes; does not specify or repair stack-safety properties \\
\midrule

\textbf{Arbiter}~\citep{vadayath:22} &
Multiple binary vulnerability classes &
Hybrid static and dynamic analysis around vulnerable sinks &
Detects several bug classes in binaries &
Detection only; no stack-specific model checking or automated repair \\
\midrule

\textbf{BoKASAN} / \textbf{MTSan}~\citep{BoKASAN,mtsan} &
Runtime memory-safety errors in binaries &
Binary-only sanitisation, object recovery, and memory-safety instrumentation &
Provides strong runtime bug detection during testing and fuzzing &
Coverage-dependent runtime checking; no formal stack-property verification or repair \\
\midrule

\textbf{Operation Mango}~\citep{mango} &
Firmware taint-style vulnerabilities &
Static binary data-flow analysis &
Scales taint analysis to firmware binaries &
Targets taint propagation rather than stack-state verification; no patching \\
\midrule

\textbf{HeapHopper}~\citep{heaphopper} &
Exploit discovery in heap implementations &
Bounded model checking and symbolic execution over heap-operation sequences &
Automatically discovers heap exploitation primitives &
Focused on heap allocators rather than stack BOs or binary repair \\
\bottomrule
\end{tabularx}
\end{table*}

\subsection{Vulnerability Discovery in Source and Binary Code}

Buffer-overflow detection has been extensively studied at the source-code level,
where analyses can exploit variable declarations, type information, array
bounds, and source-level control-flow structure. For example,
CorCA~\citep{CorCA} combines static slicing with dynamic testing to detect BOs
in C programs. Delta Pointers~\citep{delta} encode bounds information into
pointer representations and raise a fault when an out-of-bounds pointer is
dereferenced. Lhee et al.~\citep{type_assisted_bo_detection} extend the C
compiler with explicit type information about buffers, enabling runtime
detection of invalid memory operations. These approaches demonstrate the value
of combining program analysis with runtime evidence, but they assume access to
source code or compiler instrumentation.

BASICS targets the harder binary-only setting, where compilation removes much of
this structure. Binary vulnerability discovery must recover or approximate
control flow, data flow, memory objects, and calling behaviour from machine code.
Several systems address this problem for different vulnerability classes and
analysis goals. Arbiter~\citep{vadayath:22} combines static and
dynamic analysis to detect multiple classes of binary vulnerabilities. Vyper
~\citep{boudjema_vyper_2020} uses concolic execution and analysis of sensitive
memory zones for multiclass vulnerability detection. IntScope
~\citep{wang_intscope_nodate} lifts x86 binaries to an intermediate
representation and combines taint analysis with symbolic execution to detect
integer overflows, while other work uses pattern matching and dynamic symbolic
execution for the same class of bugs~\citep{staticIntOverflowBinary}. Machine
learning has also been applied to binary vulnerability discovery: VulHawk
~\citep{vulhawk} and related neural approaches~\citep{deeplearningbinaries}
embed disassembled code and train models to identify vulnerable binaries or
functions.

Recent binary-analysis systems further emphasise scalability, execution speed,
or runtime bug oracles. cwe\_checker~\citep{cwe_checker} performs static checks
for several CWE classes over lifted binary representations. ZAFL
~\citep{zfal} improves binary-only coverage-guided fuzzing through
compiler-style binary instrumentation. SymQEMU~\citep{symqemu} and
SymFit~\citep{SymFit} improve the performance of binary symbolic and
concolic execution. BoKASAN~\citep{BoKASAN} and MTSan
~\citep{mtsan} provide binary-only runtime memory-safety oracles through sanitisation, object recovery, and memory-safety instrumentation. Operation
Mango~\citep{mango} scales static binary data-flow analysis to
firmware taint-style vulnerabilities.

Table~\ref{tab:binary-tools-comparison} summarises how these systems compare to
BASICS. Existing binary tools are strong at triage, fuzzing, symbolic execution,
runtime checking, taint analysis, or exploit discovery. BASICS targets a
narrower problem: stack BOs in x86-64 binaries. Its main distinction is that it
connects an explicit stack-safety violation to a binary patch and validation
step.

\subsection{Formal Analysis of Software Security Properties}

Formal methods have been applied to software security through model checking,
abstract interpretation, symbolic execution, and temporal-property checking.
Source-level verifiers such as SLAM~\citep{slam}, BLAST~\citep{blast},
CPAchecker~\citep{CPAchecker}, CBMC~\citep{cbmc}, and
SeaHorn~\citep{seahorn} check reachability, assertion safety, pointer safety,
and related safety properties over C programs or compiler-level
representations. MOPS~\citep{mops} is closer to security-property checking: it
models C programs and security rules as automata and was later used to find
security flaws in large UNIX applications~\citep{model_checking_1_million_lines}.
These tools show the value of explicit specifications, but they rely on
source-level structure that is unavailable in the binary-only setting targeted
by BASICS.

At the binary and low-level memory-corruption level, fewer systems combine
formal reasoning with vulnerability discovery. HeapHopper~\citep{heaphopper}
uses bounded model checking and symbolic execution to discover exploit
primitives in heap allocator implementations. BASICS follows the same general
idea of restricting formal reasoning to a modelled memory component,
but targets stack-based BOs in x86-64 binaries. It models stack frames and
byte-level stack states, then checks temporal properties over return addresses,
saved frame pointers, canaries, adjacent stack objects, and unsafe writes.
Unlike source-level verifiers, BASICS does not assume source code; unlike
HeapHopper, it connects the detected violation to binary patching and
validation.

\subsection{Automated Repair and Binary Rewriting}

Automated vulnerability repair has mostly been studied at the source-code level.
SenX~\citep{senx} generates source-code patches from
vulnerability-specific safety properties and a vulnerability-triggering input.
CPR~\citep{cpr} uses concolic execution to co-explore the input
space and patch space, filtering out overfitting patches. Program Vulnerability
Repair via Inductive Inference~\citep{huang2019using} infers likely patch
invariants from program states and uses them to synthesise vulnerability
patches. These systems are relevant because, like BASICS, they connect
vulnerability evidence to repair. However, they assume source code and
source-level patch locations, whereas BASICS repairs the deployed binary
artefact directly.

For binary rewriting, BASICS builds on E9Patch~\citep{e9patch}, a static
rewriting framework for x86-64 binaries. E9Patch provides the low-level
mechanism for inserting trampoline code without requiring complete control-flow
recovery. BASICS supplies the analysis layer: it uses model-checking
counterexamples to identify vulnerable stack writes, selects a patch template,
inserts the patch through binary rewriting, and validates the patched binary on
concrete inputs. This makes BASICS less general than source-level repair
systems, but more suitable for settings where source code is unavailable.

\vspace{2mm}
In summary, existing research has mainly focused on detecting vulnerabilities at
the source code level or has applied formal methods mainly to verify functional
correctness rather than explicitly to identify security vulnerabilities.
Furthermore, vulnerability detection approaches have neglected the crucial
aspect of automated vulnerability remediation. Our proposed approach addresses
this research gap by employing formal verification to systematically discover
stack BOs directly in binary code and combines this detection with an automated
binary repair mechanism.

\section{Conclusion}
\label{seq:conclusions}

This paper presented \tool, a binary-level framework for detecting and
mitigating stack-based buffer overflows through bounded model checking and
concolic execution. \tool verifies LTL stack-security properties over a
Memory State Space recovered from the binary and uses counterexample traces to identify and patch vulnerable sinks.
The evaluation showed high-precision detection on controlled benchmarks,
achieving 86.4\% precision on SARD, and 86.1\% precision
on Juliet. All generated patches removed the observed vulnerable behavior on
the tested executions, while 41 out of the 46 patched binaries also preserved the
original exit code and standard output on benign and boundary inputs.
The open-source and Linux package sweeps further showed that
function-by-function analysis allows \tool to scale to larger binaries, 
while also highlighting the remaining limitations of bounded
symbolic execution and caller-context reconstruction.

\section*{Acknowledgments}
This work was partially supported by P2030 through project I2DT, ref. COMPETE2030-FEDER-00389100, an ITEA4 European project (ref. 22025), and by FCT through the LASIGE Research Unit, ref. UID/00408/2025, DOI \url{https://doi.org/10.54499/UID/00408/2025}.

\bibliographystyle{elsarticle-harv} 
\bibliography{refs}

\begin{thebibliography}{68}
\expandafter\ifx\csname natexlab\endcsname\relax\def\natexlab#1{#1}\fi
\providecommand{\url}[1]{\texttt{#1}}
\providecommand{\href}[2]{#2}
\providecommand{\path}[1]{#1}
\providecommand{\DOIprefix}{doi:}
\providecommand{\ArXivprefix}{arXiv:}
\providecommand{\URLprefix}{URL: }
\providecommand{\Pubmedprefix}{pmid:}
\providecommand{\doi}[1]{\href{http://dx.doi.org/#1}{\path{#1}}}
\providecommand{\Pubmed}[1]{\href{pmid:#1}{\path{#1}}}
\providecommand{\bibinfo}[2]{#2}
\ifx\xfnm\relax \def\xfnm[#1]{\unskip,\space#1}\fi
\bibitem[{Abadi et~al.(2005)Abadi, Budiu, Erlingsson and Ligatti}]{cfi}
\bibinfo{author}{Abadi, M.}, \bibinfo{author}{Budiu, M.},
  \bibinfo{author}{Erlingsson, U.}, \bibinfo{author}{Ligatti, J.},
  \bibinfo{year}{2005}.
\newblock \bibinfo{title}{Control-flow integrity}, in: \bibinfo{booktitle}{ACM
  CCS}.
\bibitem[{Aggarwal and Jalote(2006)}]{static_dynamic_detecting_vulnerabilities}
\bibinfo{author}{Aggarwal, A.}, \bibinfo{author}{Jalote, P.},
  \bibinfo{year}{2006}.
\newblock \bibinfo{title}{Integrating static and dynamic analysis for detecting
  vulnerabilities}, in: \bibinfo{booktitle}{COMPSAC}.
\bibitem[{Ball(1999)}]{the_concept_of_dynamic_analysis}
\bibinfo{author}{Ball, T.}, \bibinfo{year}{1999}.
\newblock \bibinfo{title}{The concept of dynamic analysis}.
\newblock \bibinfo{journal}{ACM SIGSOFT SEN} .
\bibitem[{Ball et~al.(2004)Ball, Cook, Levin and Rajamani}]{slam}
\bibinfo{author}{Ball, T.}, \bibinfo{author}{Cook, B.}, \bibinfo{author}{Levin,
  V.}, \bibinfo{author}{Rajamani, S.K.}, \bibinfo{year}{2004}.
\newblock \bibinfo{title}{Slam and static driver verifier: Technology transfer
  of formal methods inside microsoft}, in: \bibinfo{booktitle}{Integrated
  Formal Methods}, \bibinfo{publisher}{Springer Berlin Heidelberg}.
\bibitem[{Barabosch and Enkelmann(2024)}]{cwe_checker}
\bibinfo{author}{Barabosch, T.}, \bibinfo{author}{Enkelmann, N.E.},
  \bibinfo{year}{2024}.
\newblock \bibinfo{title}{Cwe\_checker}.
\newblock
  \bibinfo{howpublished}{\url{https://github.com/fkie-cad/cwe_checker}}.
\bibitem[{Beyer et~al.(2007)Beyer, Henzinger, Jhala and Majumdar}]{blast}
\bibinfo{author}{Beyer, D.}, \bibinfo{author}{Henzinger, T.A.},
  \bibinfo{author}{Jhala, R.}, \bibinfo{author}{Majumdar, R.},
  \bibinfo{year}{2007}.
\newblock \bibinfo{title}{The software model checker blast: Applications to
  software engineering}.
\newblock \bibinfo{journal}{STTT} .
\bibitem[{Beyer and Keremoglu(2009)}]{CPAchecker}
\bibinfo{author}{Beyer, D.}, \bibinfo{author}{Keremoglu, M.E.},
  \bibinfo{year}{2009}.
\newblock \bibinfo{title}{{CPAchecker}: A Tool for Configurable Software
  Verification}.
\newblock \bibinfo{type}{Technical Report}. School of Computing Science (CMPT),
  Simon Fraser University (SFU).
\bibitem[{Boudjema et~al.(2020)Boudjema, Verlan, Mokdad and
  Faure}]{boudjema_vyper_2020}
\bibinfo{author}{Boudjema, E.H.}, \bibinfo{author}{Verlan, S.},
  \bibinfo{author}{Mokdad, L.}, \bibinfo{author}{Faure, C.},
  \bibinfo{year}{2020}.
\newblock \bibinfo{title}{{VYPER}: {Vulnerability} detection in binary code}.
\newblock \bibinfo{journal}{Security and Privacy} .
\bibitem[{Chen et~al.(2004)Chen, Dean and
  Wagner}]{model_checking_1_million_lines}
\bibinfo{author}{Chen, H.}, \bibinfo{author}{Dean, D.},
  \bibinfo{author}{Wagner, D.A.}, \bibinfo{year}{2004}.
\newblock \bibinfo{title}{Model checking one million lines of c code}, in:
  \bibinfo{booktitle}{NDSS}.
\bibitem[{Chen and Wagner(2002)}]{mops}
\bibinfo{author}{Chen, H.}, \bibinfo{author}{Wagner, D.A.},
  \bibinfo{year}{2002}.
\newblock \bibinfo{title}{MOPS: An Infrastructure for Examining Security
  Properties of Software}.
\newblock \bibinfo{type}{Technical Report}. University of California at
  Berkeley.
\bibitem[{Chen et~al.(2022)Chen, Han, Yin, Zeng, Song, Lee, Yin and
  Shin}]{symsan_efficient_concolic_execution_data_flow_analysis}
\bibinfo{author}{Chen, J.}, \bibinfo{author}{Han, W.}, \bibinfo{author}{Yin,
  M.}, \bibinfo{author}{Zeng, H.}, \bibinfo{author}{Song, C.},
  \bibinfo{author}{Lee, B.}, \bibinfo{author}{Yin, H.}, \bibinfo{author}{Shin,
  I.}, \bibinfo{year}{2022}.
\newblock \bibinfo{title}{{SYMSAN}: Time and space efficient concolic execution
  via dynamic data-flow analysis}, in: \bibinfo{booktitle}{USENIX Security}.
\bibitem[{Chen et~al.(2023)Chen, Shi, Jiang, Li, Wang, Duan, Wang and
  Zhang}]{mtsan}
\bibinfo{author}{Chen, X.}, \bibinfo{author}{Shi, Y.}, \bibinfo{author}{Jiang,
  Z.}, \bibinfo{author}{Li, Y.}, \bibinfo{author}{Wang, R.},
  \bibinfo{author}{Duan, H.}, \bibinfo{author}{Wang, H.},
  \bibinfo{author}{Zhang, C.}, \bibinfo{year}{2023}.
\newblock \bibinfo{title}{{MTSan}: A feasible and practical memory sanitizer
  for fuzzing {COTS} binaries}, in: \bibinfo{booktitle}{USENIX Security}.
\bibitem[{Cho et~al.(2023)Cho, An, Jin and Kwon}]{BoKASAN}
\bibinfo{author}{Cho, M.}, \bibinfo{author}{An, D.}, \bibinfo{author}{Jin, H.},
  \bibinfo{author}{Kwon, T.}, \bibinfo{year}{2023}.
\newblock \bibinfo{title}{Bokasan: binary-only kernel address sanitizer for
  effective kernel fuzzing}, in: \bibinfo{booktitle}{USENIX Security}.
\bibitem[{Clarke et~al.(2004)Clarke, Kroening and Lerda}]{cbmc}
\bibinfo{author}{Clarke, E.}, \bibinfo{author}{Kroening, D.},
  \bibinfo{author}{Lerda, F.}, \bibinfo{year}{2004}.
\newblock \bibinfo{title}{A tool for checking {ANSI-C} programs}, in:
  \bibinfo{booktitle}{TACAS}.
\bibitem[{Clarke et~al.(2018)Clarke, Henzinger, Veith and
  Bloem}]{handbook_model_checking}
\bibinfo{author}{Clarke, E.M.}, \bibinfo{author}{Henzinger, T.A.},
  \bibinfo{author}{Veith, H.}, \bibinfo{author}{Bloem, R.},
  \bibinfo{year}{2018}.
\newblock \bibinfo{title}{Handbook of Model Checking}.
\newblock \bibinfo{publisher}{Springer Publishing Company, Inc.}
\bibitem[{Cowan et~al.(1998)Cowan, Pu, Maier, Hintony, Walpole, Bakke, Beattie,
  Grier, Wagle and Zhang}]{Stackguard}
\bibinfo{author}{Cowan, C.}, \bibinfo{author}{Pu, C.}, \bibinfo{author}{Maier,
  D.}, \bibinfo{author}{Hintony, H.}, \bibinfo{author}{Walpole, J.},
  \bibinfo{author}{Bakke, P.}, \bibinfo{author}{Beattie, S.},
  \bibinfo{author}{Grier, A.}, \bibinfo{author}{Wagle, P.},
  \bibinfo{author}{Zhang, Q.}, \bibinfo{year}{1998}.
\newblock \bibinfo{title}{Stackguard: automatic adaptive detection and
  prevention of buffer-overflow attacks}, in: \bibinfo{booktitle}{USENIX
  Security}.
\bibitem[{CybersecurityDive(2025)}]{cyberdive:25}
\bibinfo{author}{CybersecurityDive}, \bibinfo{year}{2025}.
\newblock \bibinfo{title}{{FBI, CISA warn hackers abusing buffer overflow CVEs
  to launch attacks}}.
\newblock
  \bibinfo{howpublished}{\url{https://www.cybersecuritydive.com/news/fbi-cisa--hackers-buffer-overflow/740072/}}.
\bibitem[{De~Moura and Bj{\o}rner(2008)}]{de2008z3}
\bibinfo{author}{De~Moura, L.}, \bibinfo{author}{Bj{\o}rner, N.},
  \bibinfo{year}{2008}.
\newblock \bibinfo{title}{Z3: An efficient smt solver}, in:
  \bibinfo{booktitle}{TACAS}.
\bibitem[{{Debian Project}()}]{debian-policy-binary-packages}
\bibinfo{author}{{Debian Project}}, .
\newblock \bibinfo{title}{{Debian Policy Manual}: Binary packages}.
\newblock \bibinfo{organization}{Debian Project}.
\newblock \URLprefix
  \url{https://www.debian.org/doc/debian-policy/ch-binary.html}.
  \bibinfo{note}{accessed 2026-05-23}.
\bibitem[{Department(2025)}]{CWETop25}
\bibinfo{author}{Department, T.U.}, \bibinfo{year}{2025}.
\newblock \bibinfo{title}{{CWE} - 2025 {CWE} top 25 most dangerous software
  weaknesses}.
\newblock
  \bibinfo{howpublished}{\url{https://cwe.mitre.org/top25/archive/2025/2025_cwe_top25.html}}.
\bibitem[{Duan et~al.(2019)Duan, Bijlani, Ji, Alrawi, Xiong, Ike,
  Saltaformaggio and Lee}]{duan2019automating}
\bibinfo{author}{Duan, R.}, \bibinfo{author}{Bijlani, A.}, \bibinfo{author}{Ji,
  Y.}, \bibinfo{author}{Alrawi, O.}, \bibinfo{author}{Xiong, Y.},
  \bibinfo{author}{Ike, M.}, \bibinfo{author}{Saltaformaggio, B.},
  \bibinfo{author}{Lee, W.}, \bibinfo{year}{2019}.
\newblock \bibinfo{title}{Automating patching of vulnerable open-source
  software versions in application binaries.}, in: \bibinfo{booktitle}{NDSS}.
\bibitem[{Duck et~al.(2020)Duck, Gao and Roychoudhury}]{e9patch}
\bibinfo{author}{Duck, G.J.}, \bibinfo{author}{Gao, X.},
  \bibinfo{author}{Roychoudhury, A.}, \bibinfo{year}{2020}.
\newblock \bibinfo{title}{Binary rewriting without control flow recovery}, in:
  \bibinfo{booktitle}{PLDI}.
\bibitem[{Duret-Lutz et~al.(2022)Duret-Lutz, Renault, Colange, Renkin, Aisse,
  Schlehuber-Caissier, Medioni, Martin, Dubois, Gillard and Lauko}]{Spot}
\bibinfo{author}{Duret-Lutz, A.}, \bibinfo{author}{Renault, E.},
  \bibinfo{author}{Colange, M.}, \bibinfo{author}{Renkin, F.},
  \bibinfo{author}{Aisse, A.G.}, \bibinfo{author}{Schlehuber-Caissier, P.},
  \bibinfo{author}{Medioni, T.}, \bibinfo{author}{Martin, A.},
  \bibinfo{author}{Dubois, J.}, \bibinfo{author}{Gillard, C.},
  \bibinfo{author}{Lauko, H.}, \bibinfo{year}{2022}.
\newblock \bibinfo{title}{From {S}pot 2.0 to {S}pot 2.10: What's new?}, in:
  \bibinfo{booktitle}{CAV}.
\bibitem[{Eckert et~al.(2018)Eckert, Bianchi, Wang, Shoshitaishvili, Kruegel
  and Vigna}]{heaphopper}
\bibinfo{author}{Eckert, M.}, \bibinfo{author}{Bianchi, A.},
  \bibinfo{author}{Wang, R.}, \bibinfo{author}{Shoshitaishvili, Y.},
  \bibinfo{author}{Kruegel, C.}, \bibinfo{author}{Vigna, G.},
  \bibinfo{year}{2018}.
\newblock \bibinfo{title}{{HeapHopper}: Bringing bounded model checking to heap
  implementation security}, in: \bibinfo{booktitle}{USENIX Security}.
\bibitem[{Ferreirinha and Medeiros(2026)}]{basics_github_a}
\bibinfo{author}{Ferreirinha, L.}, \bibinfo{author}{Medeiros, I.},
  \bibinfo{year}{2026}.
\newblock \bibinfo{title}{{BASICS}}.
\newblock \bibinfo{howpublished}{\url{https://github.com/Singularitty/BASICS}}.
\bibitem[{Gerth et~al.(1995)Gerth, Dolech, Peled, Vardi and
  Wolper}]{simple_verification_ltl}
\bibinfo{author}{Gerth, R.}, \bibinfo{author}{Dolech, D.},
  \bibinfo{author}{Peled, D.}, \bibinfo{author}{Vardi, M.},
  \bibinfo{author}{Wolper, P.}, \bibinfo{year}{1995}.
\newblock \bibinfo{title}{Simple on-the-fly automatic verification of linear
  temporal logic}.
\newblock \bibinfo{journal}{LICS} .
\bibitem[{Gibbs et~al.(2024)Gibbs, Raj, Vadayath, Tay, Miller, Ajayan, Basque,
  Dutcher, Dong, Maso, Vigna, Kruegel, Doup\'{e}, Shoshitaishvili and
  Wang}]{mango}
\bibinfo{author}{Gibbs, W.}, \bibinfo{author}{Raj, A.S.},
  \bibinfo{author}{Vadayath, J.M.}, \bibinfo{author}{Tay, H.J.},
  \bibinfo{author}{Miller, J.}, \bibinfo{author}{Ajayan, A.},
  \bibinfo{author}{Basque, Z.L.}, \bibinfo{author}{Dutcher, A.},
  \bibinfo{author}{Dong, F.}, \bibinfo{author}{Maso, X.},
  \bibinfo{author}{Vigna, G.}, \bibinfo{author}{Kruegel, C.},
  \bibinfo{author}{Doup\'{e}, A.}, \bibinfo{author}{Shoshitaishvili, Y.},
  \bibinfo{author}{Wang, R.}, \bibinfo{year}{2024}.
\newblock \bibinfo{title}{Operation mango: scalable discovery of taint-style
  vulnerabilities in binary firmware services}, in: \bibinfo{booktitle}{USENIX
  Security}.
\bibitem[{GNU and Foundation(1986)}]{GNU1986}
\bibinfo{author}{GNU}, \bibinfo{author}{Foundation, F.S.},
  \bibinfo{year}{1986}.
\newblock \bibinfo{title}{{GDB}: The gnu project debugger}.
\newblock \bibinfo{note}{Https://sourceware.org/gdb/}.
\bibitem[{Gurfinkel et~al.(2015)Gurfinkel, Kahsai and Navas}]{seahorn}
\bibinfo{author}{Gurfinkel, A.}, \bibinfo{author}{Kahsai, T.},
  \bibinfo{author}{Navas, J.A.}, \bibinfo{year}{2015}.
\newblock \bibinfo{title}{Seahorn: A framework for verifying c programs
  competition contribution}, in: \bibinfo{booktitle}{TACAS}.
\bibitem[{Holzmann(1997)}]{spin_paper}
\bibinfo{author}{Holzmann, G.}, \bibinfo{year}{1997}.
\newblock \bibinfo{title}{The model checker spin}.
\newblock \bibinfo{journal}{IEEE TSE} .
\bibitem[{Hu et~al.(2019)Hu, Zhang and Gu}]{patching_vulnerabilities}
\bibinfo{author}{Hu, Y.}, \bibinfo{author}{Zhang, Y.}, \bibinfo{author}{Gu,
  D.}, \bibinfo{year}{2019}.
\newblock \bibinfo{title}{Automatically patching vulnerabilities of binary
  programs via code transfer from correct versions}.
\newblock \bibinfo{journal}{IEEE Access} .
\bibitem[{Huang et~al.(2019)Huang, Lie, Tan and Jaeger}]{senx}
\bibinfo{author}{Huang, Z.}, \bibinfo{author}{Lie, D.}, \bibinfo{author}{Tan,
  G.}, \bibinfo{author}{Jaeger, T.}, \bibinfo{year}{2019}.
\newblock \bibinfo{title}{Using safety properties to generate vulnerability
  patches.}, in: \bibinfo{booktitle}{IEEE S\&P}.
\bibitem[{In{\'a}cio and Medeiros(2023)}]{CorCA}
\bibinfo{author}{In{\'a}cio, J.}, \bibinfo{author}{Medeiros, I.},
  \bibinfo{year}{2023}.
\newblock \bibinfo{title}{{CorCA: An Automatic Program Repair Tool for Checking
  and Removing Effectively C Flaws}}.
\newblock \bibinfo{journal}{ICST} .
\bibitem[{Keenlab(2022)}]{binabsinspector}
\bibinfo{author}{Keenlab}, \bibinfo{year}{2022}.
\newblock \bibinfo{title}{Binabsinspector}.
\newblock
  \bibinfo{howpublished}{\url{https://github.com/KeenSecurityLab/BinAbsInspector}}.
\bibitem[{Kim et~al.(2015)Kim, Kim and
  Park}]{vulnerability_detection_static_dynamic_analysis}
\bibinfo{author}{Kim, S.}, \bibinfo{author}{Kim, R.Y.C.},
  \bibinfo{author}{Park, Y.B.}, \bibinfo{year}{2015}.
\newblock \bibinfo{title}{Software vulnerability detection methodology combined
  with static and dynamic analysis}.
\newblock \bibinfo{journal}{Wireless Pers. Commun.} .
\bibitem[{Kroes et~al.(2018)Kroes, Koning, Van Der~Kouwe, Bos and
  Giuffrida}]{delta}
\bibinfo{author}{Kroes, T.}, \bibinfo{author}{Koning, K.}, \bibinfo{author}{Van
  Der~Kouwe, E.}, \bibinfo{author}{Bos, H.}, \bibinfo{author}{Giuffrida, C.},
  \bibinfo{year}{2018}.
\newblock \bibinfo{title}{Delta pointers: buffer overflow checks without the
  checks}, in: \bibinfo{booktitle}{EuroSys}.
\bibitem[{Lhee and Chapin(2002)}]{type_assisted_bo_detection}
\bibinfo{author}{Lhee, K.s.}, \bibinfo{author}{Chapin, S.J.},
  \bibinfo{year}{2002}.
\newblock \bibinfo{title}{Type-assisted dynamic buffer overflow detection}, in:
  \bibinfo{booktitle}{USENIX Security}.
\bibitem[{Lin et~al.(2007)Lin, Jiang, Xu, Mao and
  Xie}]{autopag_automated_software_patch_generation}
\bibinfo{author}{Lin, Z.}, \bibinfo{author}{Jiang, X.}, \bibinfo{author}{Xu,
  D.}, \bibinfo{author}{Mao, B.}, \bibinfo{author}{Xie, L.},
  \bibinfo{year}{2007}.
\newblock \bibinfo{title}{{AutoPaG}: towards automated software patch
  generation with source code root cause identification and repair}, in:
  \bibinfo{booktitle}{ASIACCS}.
\bibitem[{Liu et~al.(2019)Liu, Silvestro, Wang, Duan and Liu}]{liu_csod_2019}
\bibinfo{author}{Liu, H.}, \bibinfo{author}{Silvestro, S.},
  \bibinfo{author}{Wang, X.}, \bibinfo{author}{Duan, L.}, \bibinfo{author}{Liu,
  T.}, \bibinfo{year}{2019}.
\newblock \bibinfo{title}{{CSOD}: {Context}-{Sensitive} {Overflow}
  {Detection}}, in: \bibinfo{booktitle}{CGO}.
\bibitem[{Luo et~al.(2023)Luo, Wang, Wang, Tang, Xie, Zhou, Liu and
  Lu}]{vulhawk}
\bibinfo{author}{Luo, Z.}, \bibinfo{author}{Wang, P.}, \bibinfo{author}{Wang,
  B.}, \bibinfo{author}{Tang, Y.}, \bibinfo{author}{Xie, W.},
  \bibinfo{author}{Zhou, X.}, \bibinfo{author}{Liu, D.}, \bibinfo{author}{Lu,
  K.}, \bibinfo{year}{2023}.
\newblock \bibinfo{title}{Vulhawk: Cross-architecture vulnerability detection
  with entropy-based binary code search}, in: \bibinfo{booktitle}{NDSS}.
\bibitem[{Meng and Miller(2016)}]{binary_code_is_not_easy}
\bibinfo{author}{Meng, X.}, \bibinfo{author}{Miller, B.P.},
  \bibinfo{year}{2016}.
\newblock \bibinfo{title}{Binary code is not easy}, in:
  \bibinfo{booktitle}{ISSTA}.
\bibitem[{{MITRE Corporation - Common Weakness Enumeration}(2025a)}]{cwe121}
\bibinfo{author}{{MITRE Corporation - Common Weakness Enumeration}},
  \bibinfo{year}{2025}a.
\newblock \bibinfo{title}{{CWE-121}: Stack-based buffer overflow}.
\newblock
  \bibinfo{howpublished}{\url{https://cwe.mitre.org/data/definitions/121.html}}.
\bibitem[{{MITRE Corporation - Common Weakness Enumeration}(2025b)}]{cwe787}
\bibinfo{author}{{MITRE Corporation - Common Weakness Enumeration}},
  \bibinfo{year}{2025}b.
\newblock \bibinfo{title}{{CWE-787}: Out-of-bounds write}.
\newblock
  \bibinfo{howpublished}{\url{https://cwe.mitre.org/data/definitions/787.html}}.
\bibitem[{Monperrus(2018)}]{monperrus_repair}
\bibinfo{author}{Monperrus, M.}, \bibinfo{year}{2018}.
\newblock \bibinfo{title}{Automatic software repair: A bibliography}.
\newblock \bibinfo{journal}{ACM Comput. Surv.} .
\bibitem[{Mossberg et~al.(2019)Mossberg, Manzano, Hennenfent, Groce, Grieco,
  Feist, Brunson and Dinaburg}]{manticore}
\bibinfo{author}{Mossberg, M.}, \bibinfo{author}{Manzano, F.},
  \bibinfo{author}{Hennenfent, E.}, \bibinfo{author}{Groce, A.},
  \bibinfo{author}{Grieco, G.}, \bibinfo{author}{Feist, J.},
  \bibinfo{author}{Brunson, T.}, \bibinfo{author}{Dinaburg, A.},
  \bibinfo{year}{2019}.
\newblock \bibinfo{title}{Manticore: A user-friendly symbolic execution
  framework for binaries and smart contracts}.
\bibitem[{Mouzarani et~al.(2017)Mouzarani, Sadeghiyan and
  Zolfaghari}]{injection_vulns_concolic}
\bibinfo{author}{Mouzarani, M.}, \bibinfo{author}{Sadeghiyan, B.},
  \bibinfo{author}{Zolfaghari, M.}, \bibinfo{year}{2017}.
\newblock \bibinfo{title}{Detecting injection vulnerabilities in executable
  codes with concolic execution}, in: \bibinfo{booktitle}{ICSESS}.
\bibitem[{Nagy et~al.(2021)Nagy, Nguyen-Tuong, Hiser, Davidson and
  Hicks}]{zfal}
\bibinfo{author}{Nagy, S.}, \bibinfo{author}{Nguyen-Tuong, A.},
  \bibinfo{author}{Hiser, J.D.}, \bibinfo{author}{Davidson, J.W.},
  \bibinfo{author}{Hicks, M.}, \bibinfo{year}{2021}.
\newblock \bibinfo{title}{Breaking through binaries: Compiler-quality
  instrumentation for better binary-only fuzzing}, in:
  \bibinfo{booktitle}{USENIX Security}.
\bibitem[{NIST(2017)}]{nist-juliet}
\bibinfo{author}{NIST}, \bibinfo{year}{2017}.
\newblock \bibinfo{title}{Juliet c/c++ 1.3}.
\newblock
  \bibinfo{howpublished}{\url{https://samate.nist.gov/SARD/test-suites/112}}.
\bibitem[{NIST(2021)}]{nist-sard}
\bibinfo{author}{NIST}, \bibinfo{year}{2021}.
\newblock \bibinfo{title}{Software assurance reference dataset (sard)}.
\newblock
  \bibinfo{howpublished}{\url{https://www.nist.gov/itl/ssd/software-quality-group/samate/software-assurance-reference-dataset-sard}}.
\bibitem[{{Open Source Security Foundation}(2026)}]{openssf-compiler-hardening}
\bibinfo{author}{{Open Source Security Foundation}}, \bibinfo{year}{2026}.
\newblock \bibinfo{title}{{Compiler Options Hardening Guide for C and C++}}.
\newblock \bibinfo{organization}{{Open Source Security Foundation}}.
\newblock \URLprefix
  \url{https://best.openssf.org/Compiler-Hardening-Guides/Compiler-Options-Hardening-Guide-for-C-and-C%2B%2B.html}.
\bibitem[{Poeplau and Francillon(2021)}]{symqemu}
\bibinfo{author}{Poeplau, S.}, \bibinfo{author}{Francillon, A.},
  \bibinfo{year}{2021}.
\newblock \bibinfo{title}{{SymQEMU}: Compilation-based symbolic execution for
  binaries}, in: \bibinfo{booktitle}{NDSS}.
\bibitem[{Qi et~al.(2024)Qi, Hu, Xiao and Yin}]{SymFit}
\bibinfo{author}{Qi, Z.}, \bibinfo{author}{Hu, J.}, \bibinfo{author}{Xiao, Z.},
  \bibinfo{author}{Yin, H.}, \bibinfo{year}{2024}.
\newblock \bibinfo{title}{{SymFit}: Making the common (concrete) case fast for
  {Binary-Code} concolic execution}, in: \bibinfo{booktitle}{USENIX Security}.
\bibitem[{Russell et~al.(2018)Russell, Kim, Hamilton, Lazovich, Harer, Ozdemir,
  Ellingwood and McConley}]{russell2018automated}
\bibinfo{author}{Russell, R.L.}, \bibinfo{author}{Kim, L.},
  \bibinfo{author}{Hamilton, L.H.}, \bibinfo{author}{Lazovich, T.},
  \bibinfo{author}{Harer, J.A.}, \bibinfo{author}{Ozdemir, O.},
  \bibinfo{author}{Ellingwood, P.M.}, \bibinfo{author}{McConley, M.W.},
  \bibinfo{year}{2018}.
\newblock \bibinfo{title}{Automated vulnerability detection in source code
  using deep representation learning}.
\newblock \href{http://arxiv.org/abs/1807.04320}{{\tt arXiv:1807.04320}}.
\bibitem[{Schaad and Binder(2022)}]{deeplearningbinaries}
\bibinfo{author}{Schaad, A.}, \bibinfo{author}{Binder, D.},
  \bibinfo{year}{2022}.
\newblock \bibinfo{title}{Deep-learning-based vulnerability detection in binary
  executables}, in: \bibinfo{booktitle}{FPS}.
\bibitem[{Sen(2007)}]{concolic_testing}
\bibinfo{author}{Sen, K.}, \bibinfo{year}{2007}.
\newblock \bibinfo{title}{Concolic testing}, in: \bibinfo{booktitle}{ASE}.
\bibitem[{Shacham et~al.(2004)Shacham, Page, Pfaff, Goh, Modadugu and
  Boneh}]{on_the_effectiveness_address_randomization}
\bibinfo{author}{Shacham, H.}, \bibinfo{author}{Page, M.},
  \bibinfo{author}{Pfaff, B.}, \bibinfo{author}{Goh, E.J.},
  \bibinfo{author}{Modadugu, N.}, \bibinfo{author}{Boneh, D.},
  \bibinfo{year}{2004}.
\newblock \bibinfo{title}{On the effectiveness of address-space randomization},
  in: \bibinfo{booktitle}{ACM CCS}.
\bibitem[{Shariffdeen et~al.(2021)Shariffdeen, Noller, Grunske and
  Roychoudhury}]{cpr}
\bibinfo{author}{Shariffdeen, R.}, \bibinfo{author}{Noller, Y.},
  \bibinfo{author}{Grunske, L.}, \bibinfo{author}{Roychoudhury, A.},
  \bibinfo{year}{2021}.
\newblock \bibinfo{title}{Concolic program repair}, in:
  \bibinfo{booktitle}{PLDI}.
\bibitem[{Shoshitaishvili et~al.(2016)Shoshitaishvili, Wang, Salls, Stephens,
  Polino, Dutcher, Grosen, Feng, Hauser, Kruegel and Vigna}]{angr}
\bibinfo{author}{Shoshitaishvili, Y.}, \bibinfo{author}{Wang, R.},
  \bibinfo{author}{Salls, C.}, \bibinfo{author}{Stephens, N.},
  \bibinfo{author}{Polino, M.}, \bibinfo{author}{Dutcher, A.},
  \bibinfo{author}{Grosen, J.}, \bibinfo{author}{Feng, S.},
  \bibinfo{author}{Hauser, C.}, \bibinfo{author}{Kruegel, C.},
  \bibinfo{author}{Vigna, G.}, \bibinfo{year}{2016}.
\newblock \bibinfo{title}{{SoK: (State of) The Art of War: Offensive Techniques
  in Binary Analysis}}, in: \bibinfo{booktitle}{IEEE S\&P}.
\bibitem[{Szekeres et~al.(2013)Szekeres, Payer, Wei and Song}]{szekeres2013sok}
\bibinfo{author}{Szekeres, L.}, \bibinfo{author}{Payer, M.},
  \bibinfo{author}{Wei, T.}, \bibinfo{author}{Song, D.}, \bibinfo{year}{2013}.
\newblock \bibinfo{title}{Sok: Eternal war in memory}, in:
  \bibinfo{booktitle}{IEEE S\&P}.
\bibitem[{{The Linux Kernel Organization}(2025)}]{linux-kernel-userspace-api}
\bibinfo{author}{{The Linux Kernel Organization}}, \bibinfo{year}{2025}.
\newblock \bibinfo{title}{{The Linux Kernel User-Space API Guide}}.
\newblock \URLprefix \url{https://docs.kernel.org/userspace-api/index.html}.
\bibitem[{Vadayath et~al.(2022)Vadayath, Eckert, Zeng, Weideman, Menon,
  Fratantonio, Balzarotti, Doupé, Wang, Hauser and
  Shoshitaishvili}]{vadayath:22}
\bibinfo{author}{Vadayath, J.}, \bibinfo{author}{Eckert, M.},
  \bibinfo{author}{Zeng, K.}, \bibinfo{author}{Weideman, N.},
  \bibinfo{author}{Menon, G.P.}, \bibinfo{author}{Fratantonio, Y.},
  \bibinfo{author}{Balzarotti, D.}, \bibinfo{author}{Doupé, A.},
  \bibinfo{author}{Wang, R.}, \bibinfo{author}{Hauser, C.},
  \bibinfo{author}{Shoshitaishvili, Y.}, \bibinfo{year}{2022}.
\newblock \bibinfo{title}{Arbiter: {Bridging} the {Static} and {Dynamic}
  {Divide} in {Vulnerability} {Discovery} on {Binary} {Programs}}, in:
  \bibinfo{booktitle}{USENIX Security}.
\bibitem[{Wang et~al.(2009)Wang, Wei, Lin and Zou}]{wang_intscope_nodate}
\bibinfo{author}{Wang, T.}, \bibinfo{author}{Wei, T.}, \bibinfo{author}{Lin,
  Z.}, \bibinfo{author}{Zou, W.}, \bibinfo{year}{2009}.
\newblock \bibinfo{title}{{IntScope}: {Automatically} {Detecting} {Integer}
  {Overﬂow} {Vulnerability} in {X86} {Binary} {Using} {Symbolic}
  {Execution}}, in: \bibinfo{booktitle}{NDSS}.
\bibitem[{Wang et~al.(2021)Wang, Fan, Yu and Meng}]{BOFSanitizer}
\bibinfo{author}{Wang, W.}, \bibinfo{author}{Fan, M.}, \bibinfo{author}{Yu,
  A.}, \bibinfo{author}{Meng, D.}, \bibinfo{year}{2021}.
\newblock \bibinfo{title}{{BOFSanitizer: Efficient locator and detector for
  buffer overflow vulnerability}}, in:
  \bibinfo{booktitle}{HPCC/DSS/SmartCity/DependSys}.
\bibitem[{Wheeler(2025)}]{flawfider}
\bibinfo{author}{Wheeler, D.}, \bibinfo{year}{2025}.
\newblock \bibinfo{title}{Flawfinder}.
\newblock \bibinfo{howpublished}{\url{https://dwheeler.com/flawfinder/}}.
\bibitem[{Xu and Wang(2022)}]{BofAEG}
\bibinfo{author}{Xu, S.}, \bibinfo{author}{Wang, Y.}, \bibinfo{year}{2022}.
\newblock \bibinfo{title}{{BofAEG: Automated Stack Buffer Overflow
  Vulnerability Detection and Exploit Generation Based on Symbolic Execution
  and Dynamic Analysis}}.
\newblock \bibinfo{journal}{Security Commun. Netw.} .
\bibitem[{Yun et~al.(2018)Yun, Lee, Xu, Jang and
  Kim}]{qsym_concolic_execution_hybrid_fuzzing}
\bibinfo{author}{Yun, I.}, \bibinfo{author}{Lee, S.}, \bibinfo{author}{Xu, M.},
  \bibinfo{author}{Jang, Y.}, \bibinfo{author}{Kim, T.}, \bibinfo{year}{2018}.
\newblock \bibinfo{title}{{QSYM}: A practical concolic execution engine
  tailored for hybrid fuzzing}, in: \bibinfo{booktitle}{USENIX Security}.
\bibitem[{Zhang et~al.(2022)Zhang, Gao, Duck and Roychoudhury}]{huang2019using}
\bibinfo{author}{Zhang, Y.}, \bibinfo{author}{Gao, X.}, \bibinfo{author}{Duck,
  G.J.}, \bibinfo{author}{Roychoudhury, A.}, \bibinfo{year}{2022}.
\newblock \bibinfo{title}{Program vulnerability repair via inductive
  inference}, in: \bibinfo{booktitle}{ISSTA}.
\bibitem[{Zhang et~al.(2015)Zhang, Sun, Deng, Cheng, Zeng, Fu and
  Feng}]{staticIntOverflowBinary}
\bibinfo{author}{Zhang, Y.}, \bibinfo{author}{Sun, X.}, \bibinfo{author}{Deng,
  Y.}, \bibinfo{author}{Cheng, L.}, \bibinfo{author}{Zeng, S.},
  \bibinfo{author}{Fu, Y.}, \bibinfo{author}{Feng, D.}, \bibinfo{year}{2015}.
\newblock \bibinfo{title}{Improving accuracy of static integer overflow
  detection in binary}, in: \bibinfo{booktitle}{RAID}.

\end{thebibliography}

\subsection*{  } %
\setlength\intextsep{0pt} %
\begin{wrapfigure}{l}{0.13\textwidth}
    \centering
    \includegraphics[width=0.15\textwidth]{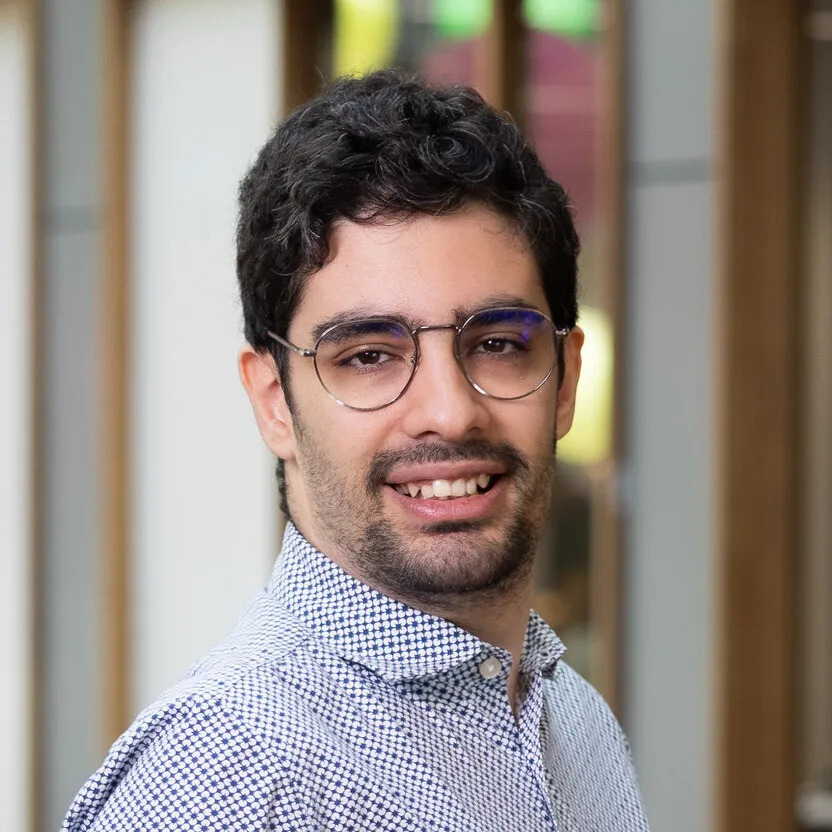}
\end{wrapfigure}
\noindent \textbf{Luís Ferreirinha} is a PhD Student at Vrije Universitiet Amsterdam, conducting research in programming languages and systems security. He holds an MSc in Informatics from the Faculty of Sciences of the University of Lisbon (FCUL). He participated in the ADMORPH project in activities related to detecting vulnerabilities in binaries. His research interests include programming languages, type systems, formal
verification, systems security, and cryptographic proofs. More information about him at \url{https://www.lpferreirinha.com/}.

\subsection*{  }
\setlength\intextsep{0pt} %
\begin{wrapfigure}{l}{0.13\textwidth}
    \centering
    \includegraphics[width=0.15\textwidth]{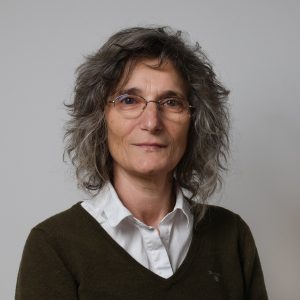}
\end{wrapfigure}
\noindent \textbf{Ib\'{e}ria Medeiros}
is an Associate Professor at the Faculty of Sciences of the University of Lisbon (FCUL), and an integrated researcher of the LASIGE research unit.
Her research interests are software security, vulnerability and attack detection, code privacy and correction, and artificial intelligence applied to cybersecurity. She is the author of tools for software security and cybersecurity, and has been involved in international and national projects. More information about her at \url{https://www.di.fc.ul.pt/~imedeiros/}.

\end{document}